\documentclass{sig-alternate-10pt}

\usepackage{cite}

\usepackage{algorithm}
\usepackage{algorithmic}
\usepackage{fixltx2e}
\usepackage{amsmath}
\usepackage{float}
\usepackage{cite}
\usepackage{url}
\usepackage{booktabs} 
\usepackage{epstopdf}

\usepackage{xr}
\usepackage{multirow}

\begin{document}

\title{Energy-Performance Trade-offs \\in Mobile Data Transfers}

\numberofauthors{2}
\author{
\alignauthor
Kemal Guner\\
       \affaddr{Computer Science and Engineering}\\
       \affaddr{University at Buffalo (SUNY)}\\
       \affaddr{Buffalo, New York 14260}\\
       \email{kemalgne@buffalo.edu}
\alignauthor
Tevfik Kosar\\
       \affaddr{Computer Science and Engineering}\\
       \affaddr{University at Buffalo (SUNY)}\\
       \affaddr{Buffalo, New York 14260}\\
       \email{tkosar@buffalo.edu}
}    

\maketitle
\begin{abstract}
By year 2020, the number of smartphone users globally will reach 3 Billion and the mobile data traffic (cellular + WiFi) will exceed PC internet traffic the first time. 
As the number of smartphone users and the amount of data transferred per smartphone grow exponentially, limited battery power is becoming an increasingly critical problem for mobile devices which increasingly depend on network I/O. Despite the growing body of research in power management techniques for the mobile devices at the hardware layer as well as the lower layers of the networking stack, there has been little work focusing on saving energy at the application layer for the mobile systems during network I/O. 
In this paper, to the best of our knowledge, we are first to provide an in depth analysis of the effects of application layer data transfer protocol parameters on the energy consumption of mobile phones. We show that significant energy savings can be achieved with application layer solutions at the mobile systems during data transfer with no or minimal performance penalty. In many cases, performance increase and energy savings can be achieved simultaneously. 
\end{abstract}

\section{Introduction}
\label{sec:introduction}

The number of smartphone users globally has already exceeded 2 Billion, and this number is expected to reach 3 Billion by 2020~\cite{edwards2016gamification}.
It is also estimated that smartphone mobile data traffic (cellular + WiFi) will reach 370 Exabytes per year by that time, exceeding PC internet traffic the first time in the history~\cite{Cisco_2016}. 
An average smartphone consumes between 300 -- 1200 milliwatts power~\cite{carroll2010analysis} depending on the type of applications it is running,   
and most of the energy in smartphone applications is spent for networked I/O. During an active data transfer, the cellular (i.e., GSM) and WiFi components of a smartphone consume more power than its CPU, RAM, and even LCD+graphics card at the highest brightness level \cite{pathak2012energy, carroll2010analysis}. Although the mobile data traffic and the amount of energy spent for it increase at a very fast pace, the battery capacities of smartphones do not increase at the same rate.

Limited  battery  power  is becoming an increasingly critical  problem  for smartphones and mobile computing, and many techniques have been proposed in the literature to overcome this at different layers. 
At the physical layer, techniques were proposed to choose appropriate modulation, coding, and transmission power control schemes to improve energy efficiency of the mobile device~\cite{cianca2001improving, schurgers2001modulation, cui2004energy, singh1998pamas, takai2001effects}.
At the media access control (MAC) layer, several new energy-efficient MAC protocol designs were proposed~\cite{woesner1998power, ye2002energy, bharghavan1994macaw, woo2001transmission, krashinsky2005minimizing, krashinsky2005minimizing, nasipuri2000mac}.
At the network layer, low-power and scalable routing algorithms were developed~\cite{xu2001geography, chang2000energy, seada2004energy, singh1998power, toh2001maximum}.
At the transport layer, traffic shaping techniques~\cite{akella2001protocols} and new transport protocols~\cite{zorzi1999tcp, kravets2000application, akella2001protocols, haas1997mobile, chandra2002application} were proposed to exploit application-specific information and reduce power utilization.  

Despite the growing body of research in power management techniques for the lower layers of the mobile networking stack, there has been little work focusing on saving network I/O (data transfer) energy at the application layer. 
The most notable work in this area are: tuning the client playback buffer size during media streaming in order to minimize the total energy spent~\cite{bertozzi2002power}; using lossless compression techniques to minimize the amount of data transferred as well as the energy consumed on wireless devices ~\cite{xu2003impact}; and joint optimization of the application layer, data link layer, and physical layer of the protocol stack using an application-oriented objective function in order to improve multimedia quality and power consumption at the same time~\cite{khan2006application}.
We believe a significant amount of network I/O energy savings can be obtained at the application layer with no or minimal performance penalty. Although lower-layer network stack approaches are an important part of the solution, application-layer power management is another key to optimizing network I/O energy efficiency in mobile computing, which has been long ignored. 

In this paper, we analyze the effects of different application layer data transfer protocol parameters (such as the number of parallel data streams per file, the level of concurrent file transfers to fill the mobile network pipes, and the I/O request size) on mobile data transfer throughput and energy consumption. 

In summary, our contributions within this paper are the following: 

\begin{itemize}

\item To the best of our knowledge, we are first to provide an in depth analysis of the effects of application layer data transfer protocol parameters on the energy consumption of mobile phones.

\item We show that significant energy savings can be achieved with application-layer solutions at the mobile systems during data transfer with no or minimal performance penalty.

\item We also show that, in many cases, performance increase and energy savings can be achieved simultaneously.

\end{itemize}

The rest of this paper is organized as follows: Section II presents background information on energy-aware tuning of application-layer data transfer protocol parameters and discusses the related work in this area; Section III provides the methodology of our analysis; Section IV presents an in-depth experimental analysis of the application-layer parameter effects on mobile data transfer performance and energy consumption; and Section V concludes the paper.

\section{Background}
\label{sec:background}

The majority of work on mobile device energy savings mostly focuses putting the devices to sleep during idle times~\cite{krashinsky2005minimizing, vallina2011erdos, schulman2010bartendr, vallina2013energy}. A recent study by Dogar et al.~\cite{Dogar2010} takes this approach to another step, and puts the device into sleep even during data transfer by exploiting the high-bandwidth wireless interface. They combine small gaps between packets into meaningful sleep intervals, thereby allowing the NIC as well as the device to doze off. Another track of study in this area focuses on switching among multiple radio interfaces in an attempt to reduce the overall power consumption of the mobile device~\cite{pering2006coolspots, correia2010challenges, balasubramanian2009energy, nika2015energy}.
These techniques are orthogonal to our application-layer protocol tuning approach and could be used together to achieve higher energy efficiency in the mobile systems.

The closest work to ours in the literature is the work by Bertozzi et al.~\cite{bertozzi2003transport}, in which they investigate the energy trade-off in mobile networking as a function of the TCP receive buffer size and show that the TCP buffering mechanisms can be exploited to significantly increase energy efficiency of the transport layer with minimum performance overheads.

In this work, we focus on the tuning of three different protocol tuning parameters:  concurrency (the level of concurrent file transfers to fill the mobile network pipes),
parallelism (the number of parallel data streams per file), and I/O request size.
 


{\em Concurrency} refers to sending multiple files simultaneously through the network using different data channels at the same time.
Most studies in this area do not take the data size and the network characteristics into consideration when setting the concurrency level~\cite{kosar04, Thesis_2005, Kosar09, JGrid_2012}. Liu et al.~\cite{R_Liu10} adapt the concurrency level based on the changes in the network traffic, but do not take into account other bottlenecks that can occur on the end systems. 

{\em Parallelism} sends different chunks of the same file using different data channels (i.e., TCP streams) at the same time 
and achieves high throughput by mimicking the behavior of individual streams and getting a higher share of the available bandwidth~\cite{R_Sivakumar00, R_Lee01, R_Balak98, R_Hacker05, R_Eggert00, R_Karrer06, R_Lu05, DADC_2008, DADC_2009, NDM_2011}. On the other hand, using too many simultaneous connections congests the network and the throughput starts dropping down. 
Predicting the optimal parallel stream number for a specific setting is a very challenging problem due to the dynamic nature of the interfering background traffic. 
Hacker et al. claimed that the total number of streams behaves like one giant stream that transfers in total capacity of each streams' achievable throughput~\cite{R_Hacker02}. However, this model only works for uncongested networks, since it accepts that packet loss ratio is stable and does not increase as the number of streams increases. 
Dinda et al. modeled the bandwidth of multiple streams as a partial second order polynomial which needs two different real-time throughput measurements to provide accurate predictions~\cite{R_Dinda05}. 

{\em I/O request size} is the size of request that application uses to perform I/O operation on storage device.
The I/O request size may have a big impact on the storage performance, and also on the end-to-end data transfer performance if the end system storage throughput is the main bottleneck. 


When used wisely, these parameters have a potential to improve the end-to-end data transfer performance at a great extent, but improper use of these parameters can also hurt the performance of the data transfers due to increased load at the end-systems and congested links in the network. For this reason, it is crucial to find the best combination for these parameters with the least intrusion and overhead to the system resource utilization and power consumption.
%

\begin{table*}[t]
\small
	\begin{centering}
		\begin{tabular}{ |r|r|r|r|r| }
			\hline
			\rule{0pt}{2.3ex}
			Producer & Google & Samsung &  Samsung  & Samsung \\
			\hline
			\rule{0pt}{2.3ex}
			Model & Nexus S & Galaxy Nexus N3 (L700) & Galaxy S4 & Galaxy S5  \\
			
			OS & Android 4.1.1 (API 16) & Android 4.3 (API 18) & Android 5.0.1 (API 21) & Android 5.0.1 (API 21)	\\
			
			CPU & 1.0 GHz Cortex-A8 & Dual-core 1.2 GHz & Quad-core 1.9 GHz Krait 300 & Quad-core 2.5 GHz Krait 400	\\
			
			Wifi & 802.11 b/g/n & 802.11 a/b/g/n & 802.11 a/b/g/n/ac & 802.11 a/b/g/n/ac	\\
			
			Storage & 16 GB	& 32 GB & 16 GB & 16 GB	\\
			
			Memory & 512 MB & 1 GB & 2 GB & 2 GB	\\
			\hline
			
		\end{tabular}
		\caption{Specifications of the mobile devices used in the experiments.} \label{tab:phonespecs}
	\end{centering}
\end{table*}

In the literature, several highly-accurate predictive models ~\cite{R_Yin11, R_Yildirim11, DISCS12, Cluster_2015} were developed which would require as few as three sampling points to provide very accurate predictions for the parallel stream number giving the highest transfer throughput for the wired networks. 
Yildirim et al. analyzed the combined effect of parallelism and concurrency on end-to-end data transfer throughput~\cite{TCC_2016}.
%
Managed File Transfer (MFT) systems were proposed which used a subset of these parameters in an effort to improve the end-to-end data transfer throughput~\cite{WORLDS_2004, ScienceCloud_2013, globusonline, Royal_2011, IGI_2012}.
Alan et al. analyzed the effects of parallelism and concurrency on end-to-end data transfer throughput versus total energy consumption in wide-area wired networks in the context of GridFTP data transfers~\cite{Alan2015, Kosar_jrnl14}. {\em None of the existing work in this area studied the effects of these three parameters on the mobile energy consumption and the performance versus energy trade-offs of tuning these parameters in this context.}

\section{Methodology}
\label{sec:methodology}

In our analysis, we have used a single-phase portable Yokogawa WT210 power meter, which provides highly accurate and fine granular power values (up to 10 readings per second) and is one of the accepted devices by the Standard Performance Evaluation Corporation (SPEC) power committee for power measurement and analysis purposes in the field~\cite{specOverview}. This power meter is used to measure the power consumption rates during the data transfers at the mobile client device. 

Prior to initiating any data transfer, we examined the base power state of each tested mobile device. To better understand the base power state, one should consider the following three power states: ``on'', ``idle'' and ``suspended''. A mobile device is considered ``on'' when any user applications are actively running. In the ``idle'' state, a mobile device is fully awake, but no user application is actively running. The ``suspended'' mobile device maintains only a low level activity of communication while no application is running. To measure the base power state for our experiments, we established a setting when the mobile device is in the 
``on'' state with the screen is also on (always at the same brightness level), any communication interface other than the one being tested (i.e., Wifi or 4G LTE) is disabled, and a minimum number of necessary applications are running in the background. This setup ensured that the base power of the tested mobile device is both low and in a balanced state throughput the experiments. 

We designed a real time test environment with four different mobile devices (as specifications presented in Table~\ref{tab:phonespecs}). We tested both WiFi and 4G LTE connections in progress of data transfers on end-systems. 
To reduce the effect of number of active users and the effect of peak/off-peak hours during the transfer of datasets, we adopted a strategy of using different time frames for each of the same experiment settings, and take the average throughput and energy consumption values. We conducted all experiments at the same location and with the same distance and interference for objective analysis of the end-system devices. 

We run initial tests for all four mobile devices at different times of the day to obtain robust base power for each. With the help of these values, the total energy consumption during data transfers is calculated as follows:

\vspace{-3mm}

\begin{equation}
\hspace{-2.8cm}	
E_{t}  = E_{b} + E_{d}
\end{equation}
\vspace{-2mm}
\begin{equation}
E_{d}  = \int_{t_{start}}^{t_{end}} (P_{max}(t) - P_{b}(t)) \cdot dt
\label{eq:power}
\end{equation}

\noindent where,
\vspace{1mm}

\indent $\bullet$ $E_{t}$: Total energy consumption of data transfer \\
\indent $\bullet$ $E_{d}$: Dynamic energy consumption of data transfer \\
\indent $\bullet$ $E_{b}$: Base energy consumption of data transfer \\
\indent $\bullet$ $P_{max}$: Total power consumption  \\
\indent $\bullet$ $P_{b}$: Base power consumption before initiating the test \\
\indent $\bullet$ $t_{start}$: Data transfer start time \\
\indent $\bullet$ $t_{end}$: Data transfer end time \\

Dynamic energy consumption $E_{d}$ in equation \ref{eq:power} is established by taking integral of  subtract values of base power of device from total instantaneous power measured by power meter per second as can be seen in Figure~\ref{fig:instantpower}. All the energy consumption results presented in the paper refer to dynamic energy consumption as stated in equation \ref{eq:power}. Since we aim to analyze the effect of application layer parameters on energy consumption, we ignored the energy consumed when the device is idle.

\begin{table*}[t]
\small
	\begin{centering}
		\begin{tabular}{ r@{\hskip 1cm}r@{\hskip 1cm}r @{\hskip 1cm}r@{\hskip 1cm}r }
			\hline
			{\bf }& &  & {\bf} & {\bf}\\
			{\bf Dataset Name} & {\bf Number of Files}  &  {\bf Ave. File Size}  & {\bf Min-Max } & {\bf Total Size}\\
			{\bf }& &  & {\bf} & {\bf}\\
			\hline
			\rule{0pt}{2.3ex}
			HTML	& 1500 	&  128 KB 	& 102 KB - 153 KB 	& 196 MB\\
			IMAGE	& 200 	&  640 KB 	& 524 KB - 786 KB 	& 128 MB	\\		
			VIDEO	& 64 	&  16.4 MB 	& 10 MB - 20 MB 	& 1124 MB	\\
			32GB	& 32 	&  1 GB 	& 1 GB - 1 GB 		& 32768 MB	\\
			3GB		& 1  	&  3 GB		& 3 GB - 3 GB 		& 3072 MB	\\
			10GB	& 1 	&  10GB 	& 10 GB - 10 GB 	& 10240 MB	\\
			\hline
		\end{tabular}
		\caption{Characteristics of the datasets used in the experiments.} \label{tab:dataset}
	\end{centering}
\end{table*}

\begin{figure}[t]
	\begin{center}
		\includegraphics[keepaspectratio=true,angle=0, width=90mm]{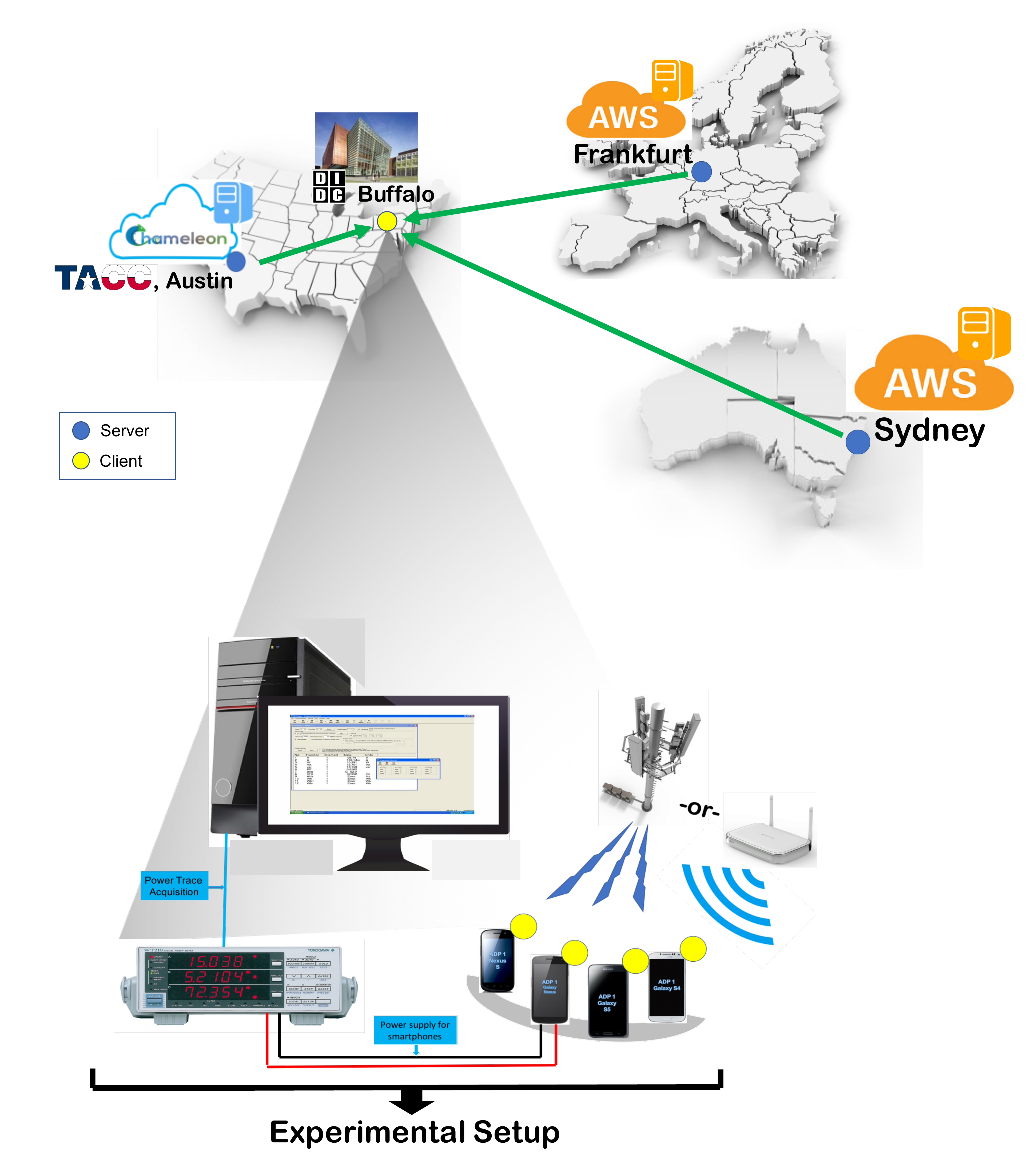}
		\vspace{-5mm}
		\caption{Network map of the experimental testbed and the setup of the power measurement system.}
		\label{fig:exSetup}
	\end{center}
	\vspace{-1cm}
\end{figure}

We choose HTTP (Hypertext Transport Protocol) as the application-layer transfer protocol to test the impact of the parameters of interest on the end-to-end data transfer throughput as well as the energy consumption of the mobile client. The main reason for this choice is that HTTP is the de-facto transport protocol for Web services ranging from file sharing to media streaming, and the studies analyzing the Internet traffic~\cite{kellerman2016daily, Mirkovic2015, Czyz2014} show that HTTP accounts for 75\% of global mobile Internet traffic.

We analyzed the data transfer throughput of HTTP data transfers and the power consumption during which we run tests with different level of concurrency (cc), parallelism (p), I/O request size, and combined concurrency \& parallelism parameters. We also measured the instantaneous power consumption and total energy consumption of each individual request among different web servers and clients.
The experiments were conducted on Amazon Elastic Compute Cloud (AWS EC2)~\cite{aws} instances, Chameleon Cloud~\cite{chameleon}, and Data Intensive Distributed Computing Laboratory (DIDCLAB) in Buffalo, NY. The 
network map of the experimental testbed and the setup of the power measurement system are illustrated in Figure~\ref{fig:exSetup}.

In the experiments, we used six different types of datasets in order to analyze the effect of each individual parameter on transfer throughput and energy consumption. The details and characteristics of these datasets are presented in Table~\ref{tab:dataset}. Overall, we generated a total dataset size of 43.5 GB where individual file sizes range between 102 KB and 10 GB. 
In order to increase the robustness of the obtained throughput and energy consumption values for each experimental setting, we run each test within the range of five to ten times, and the average values of throughput and energy consumption were used. As a result of iteration of each individual experiment among four different mobile clients and three different web servers with different bandwidth (BW) and round-trip-time (RTT), we transferred varying size of nearly 1.8 Million individual files. Due to the space limitations of the paper, we had to limit the number of graphs we can present. The detailed analysis of the application-layer parameter effects on mobile data transfer performance and energy consumption are provided and discussed in the next section.

\section{Analysis of Parameter Effects}
\label{sec:experiments}

\begin{figure*}[t]
	\begin{centering}
		\begin{tabular}{cc}
			\includegraphics[keepaspectratio=true,angle=0,width=58mm]{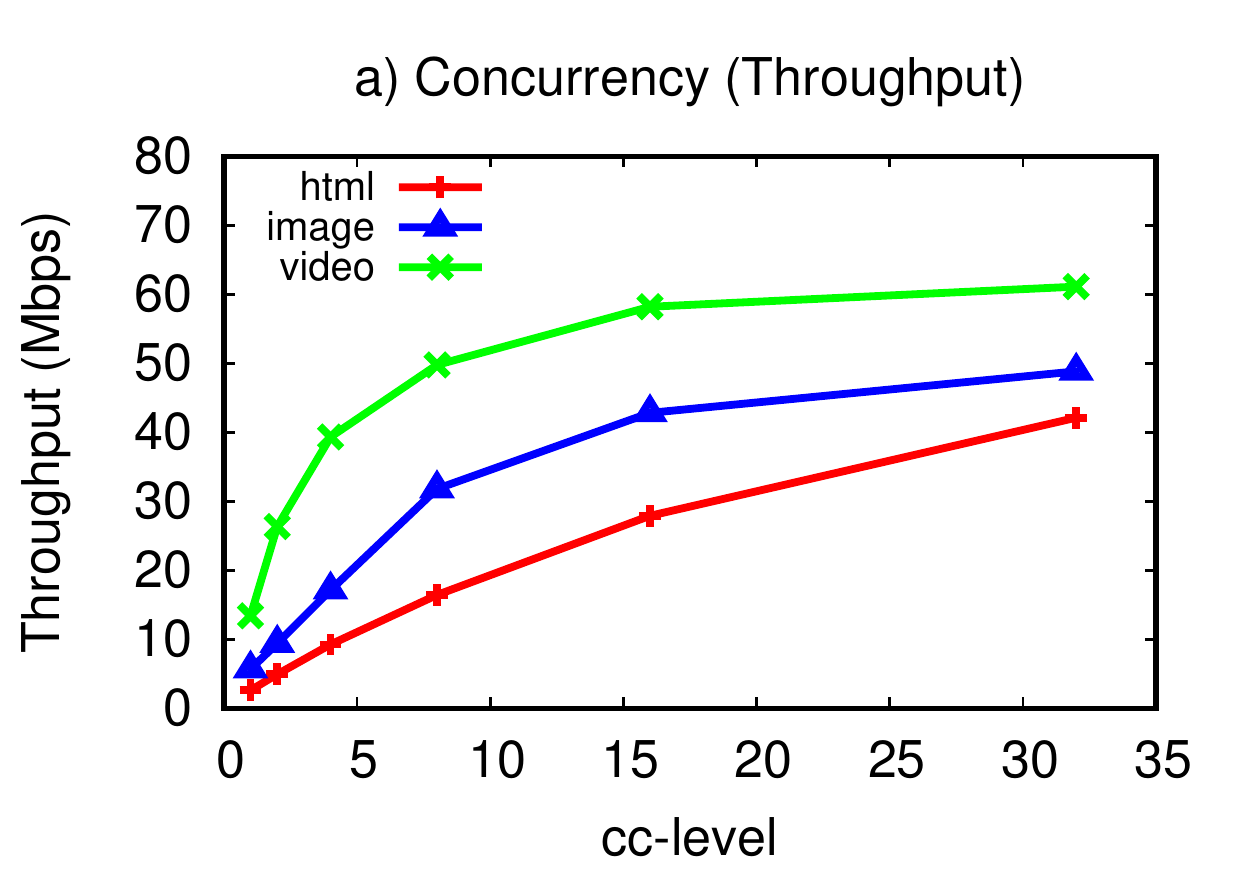}
			\includegraphics[keepaspectratio=true,angle=0,width=58mm]{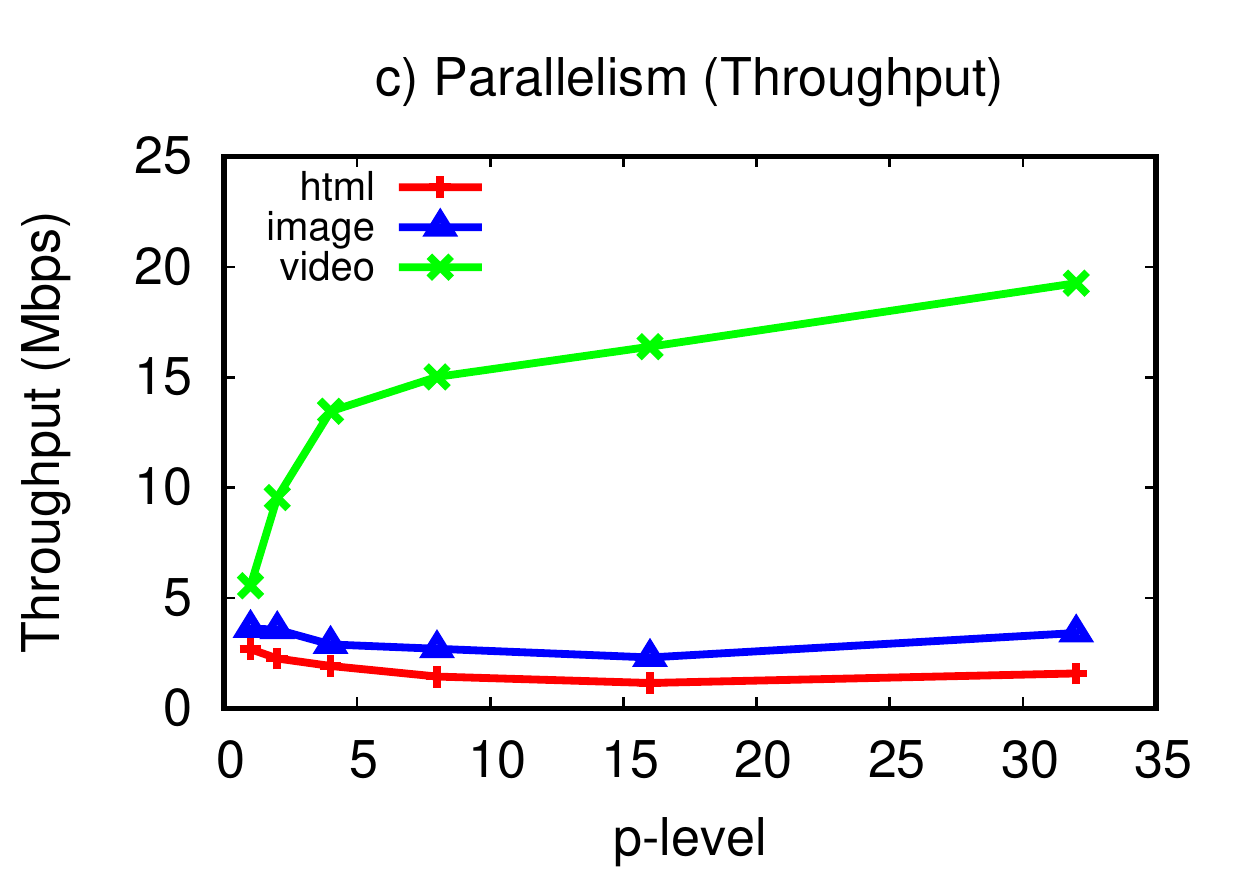}
			\includegraphics[keepaspectratio=true,angle=0,width=58mm]{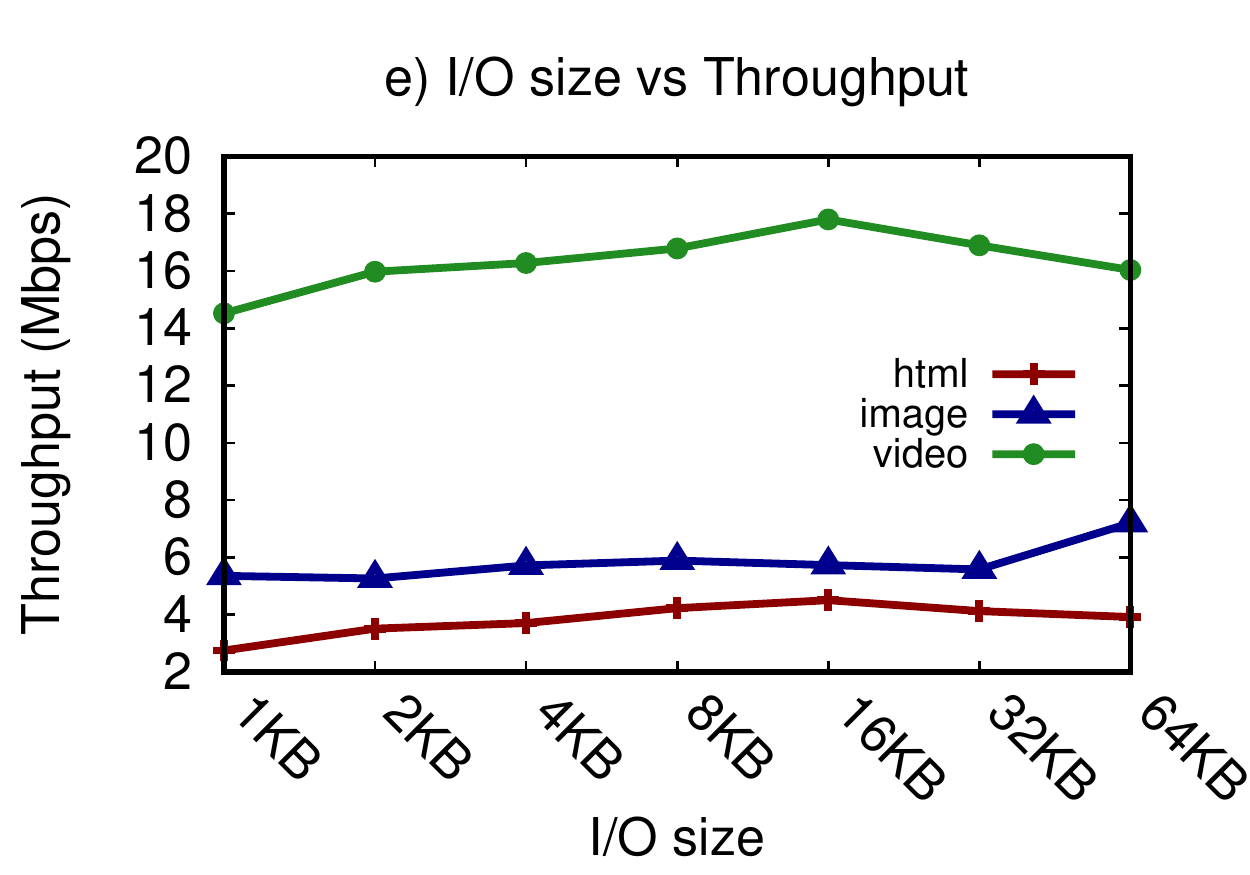}\\
			\includegraphics[keepaspectratio=true,angle=0,height=53mm,width=58mm]{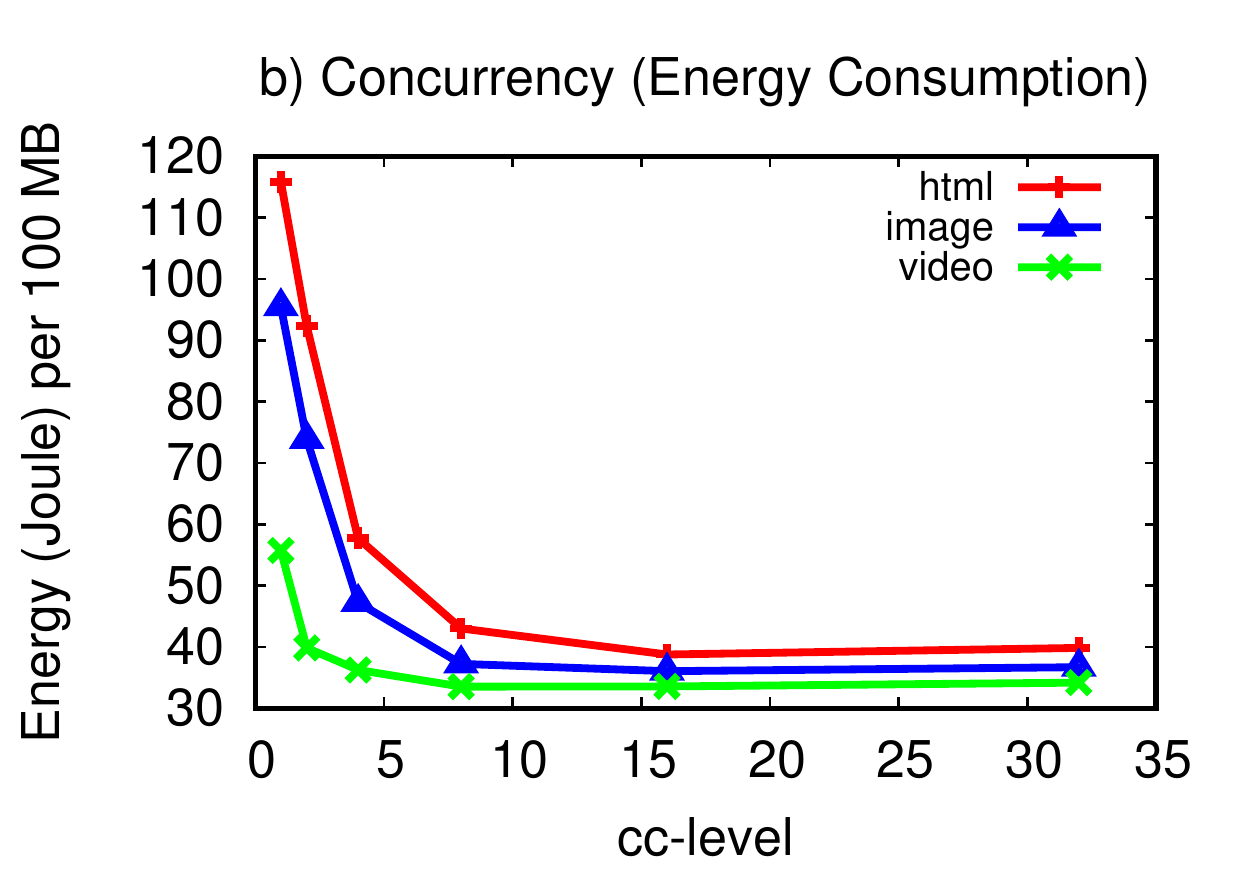}
			\includegraphics[keepaspectratio=true,angle=0,width=58mm]{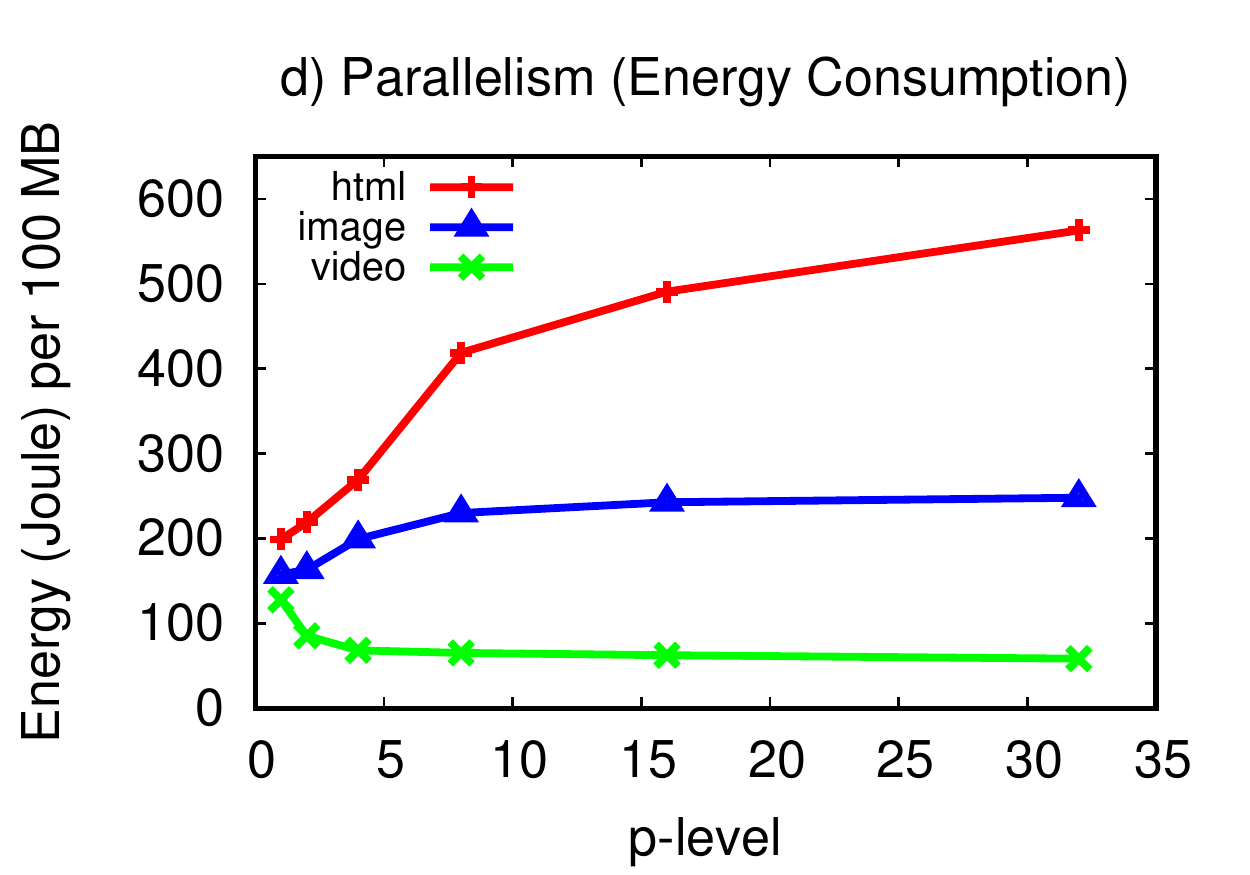}
			\includegraphics[keepaspectratio=true,angle=0,width=58mm]{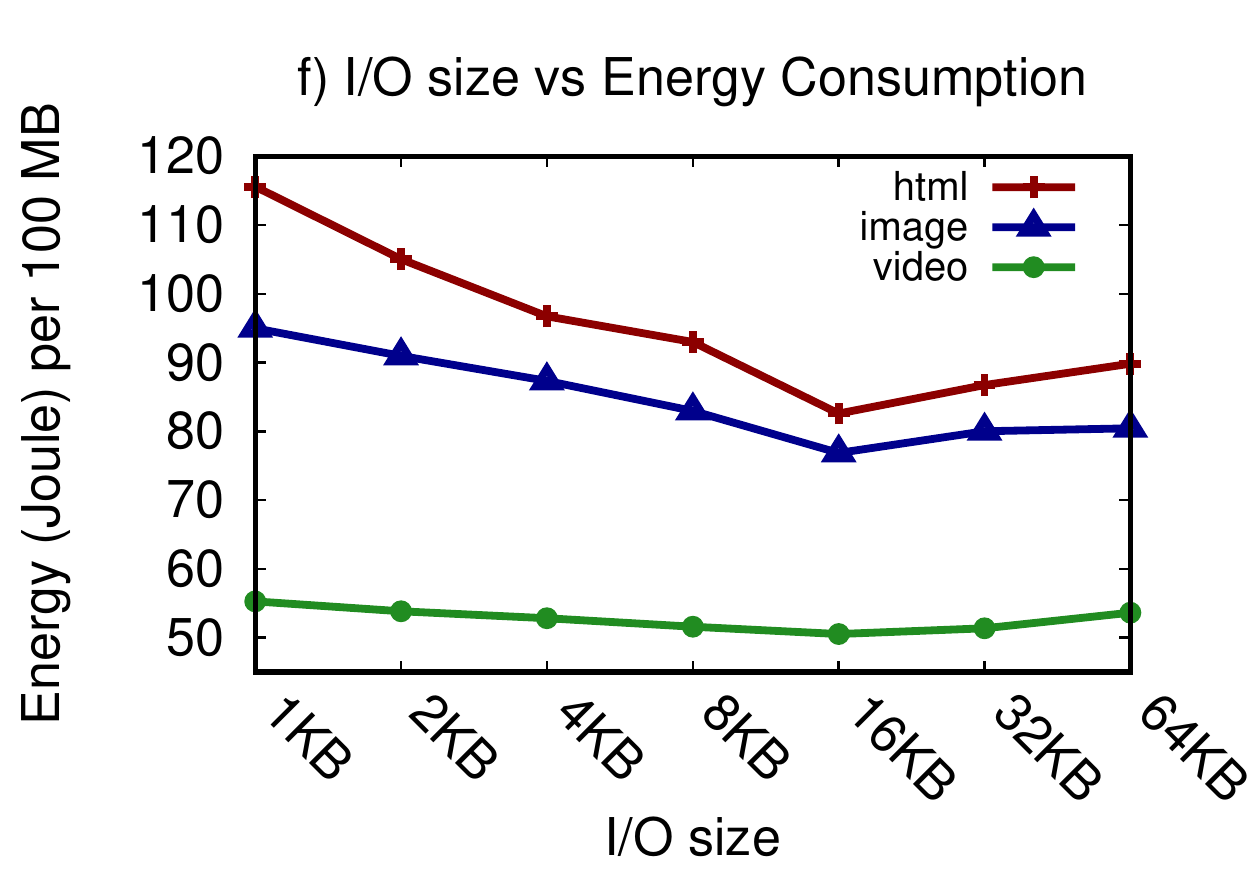}
		\end{tabular}
		\caption{Throughput vs Energy Consumption trade-offs of individual protocol parameters for WiFi data transfers between AWS EC2 Sydney and DIDCLAB Galaxy S5.} \label{fig:parameter}
	\end{centering}
\end{figure*}

One of the web servers in our testbed is located on the Chameleon Cloud at the Texas Advanced Computing Center (TACC) in Austin, Texas (USA). The other two servers are located at AWS EC2 in Sydney (Australia) and AWS EC2 in Frankfurt (Germany) respectively. These web servers are serving to the clients at DIDCLAB in Buffalo, New York (USA). The tested mobile client devices include Samsung Galaxy S5, Samsung Galaxy S4, Samsung Galaxy Nexus N3 (also known as Galaxy Nexus L700), and Google Nexus S smartphones. The web servers located at Chameleon Cloud, AWS EC2 Sydney and AWS EC2 Frankfurt are designed to support the high performance data transfer and computing within the same testbed over shared networks where the theoretical network bandwidth is 10 Gbps, 1Gbps. and 1Gbps respectively. Even though having these high bandwidth, there are two main reasons why it is hard for us to achieve such high performance during the file transfers from these web servers using mobile clients: (1) lack of dedicated network between our clients and these web servers; and (2) using limited bandwidth WiFi or cellular (4G LTE) connections. Thus, we took into account the achievable maximum throughput results of each smartphone to make a comparison on these web servers. Using a shared network connection brings other constraints such as the number of users that share the same network, data rate of provider's network and interference of other networks that use the same channel on frequency band. We used Apache HTTP web server with default configuration and custom HTTP client configuration in our data transfer experiments. All experiments were controlled on the client side and the measurements were performed using a power meter as discussed in section~\ref{sec:methodology}.

Figure~\ref{fig:parameter} presents the individual parameter effects of concurrency and parallelism on the achieved throughput and energy consumption for the the data transfers between the web server at AWS EC2 Sydney and the client Samsung Galaxy S5 at DIDCLAB in Buffalo. The energy consumption is measured per 100 MB of data transfers to make a fair comparison among different data sets. The RTT between the server and client is around 290 ms. Due to a shared network connection between the client and AWS EC2 Sydney server and limited WiFi bandwidth, the end-to-end transfer throughput of the video dataset increases from 13 Mbps up to 61 Mbps and then saturates. 

\begin{figure*}[t]
	\begin{centering}
		\begin{tabular}{cc}
			\includegraphics[keepaspectratio=true,angle=0,width=65mm]{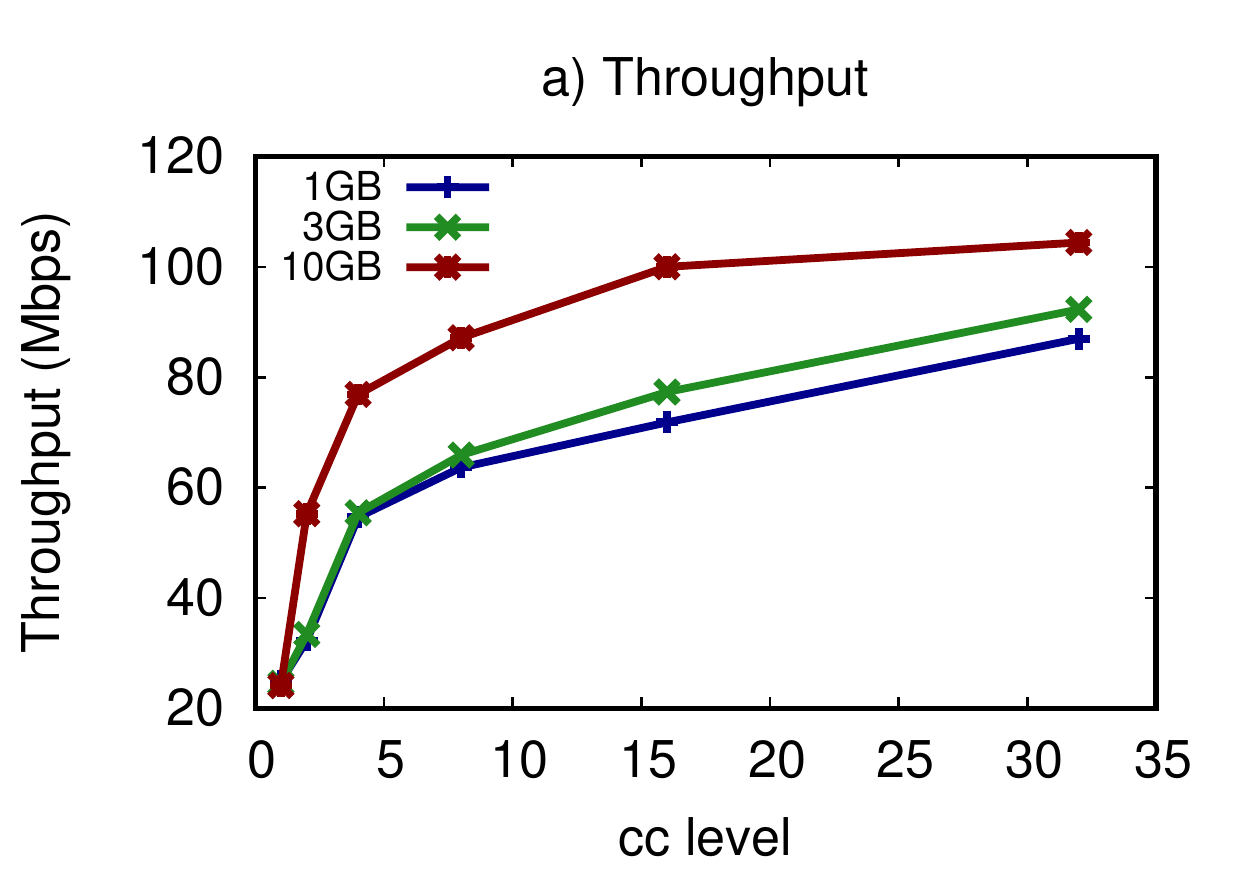}
			\hspace{1cm}
			\includegraphics[keepaspectratio=true,angle=0,width=65mm]{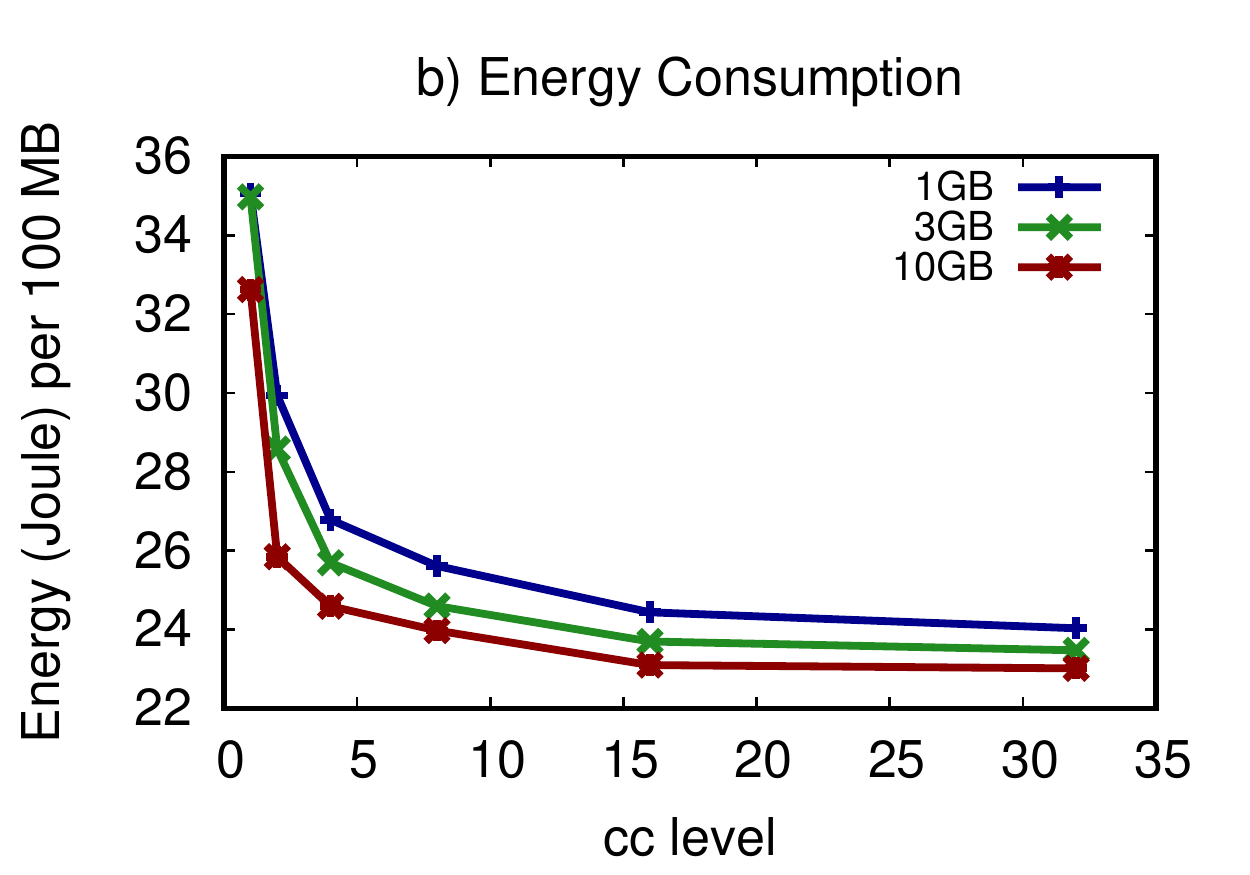}
		\end{tabular}
		\caption{A case study: Throughput vs Energy Consumption analysis of the parallelism parameter on large file transfers between AWS EC2 Sydney and DIDCLAB Galaxy S5.} \label{fig:th-pw10G}
	\end{centering}
	\vspace{-5mm}
\end{figure*}

\begin{figure*}[t]
	\begin{centering}
		\begin{tabular}{cc}
			\includegraphics[keepaspectratio=true,angle=0,width=65mm]{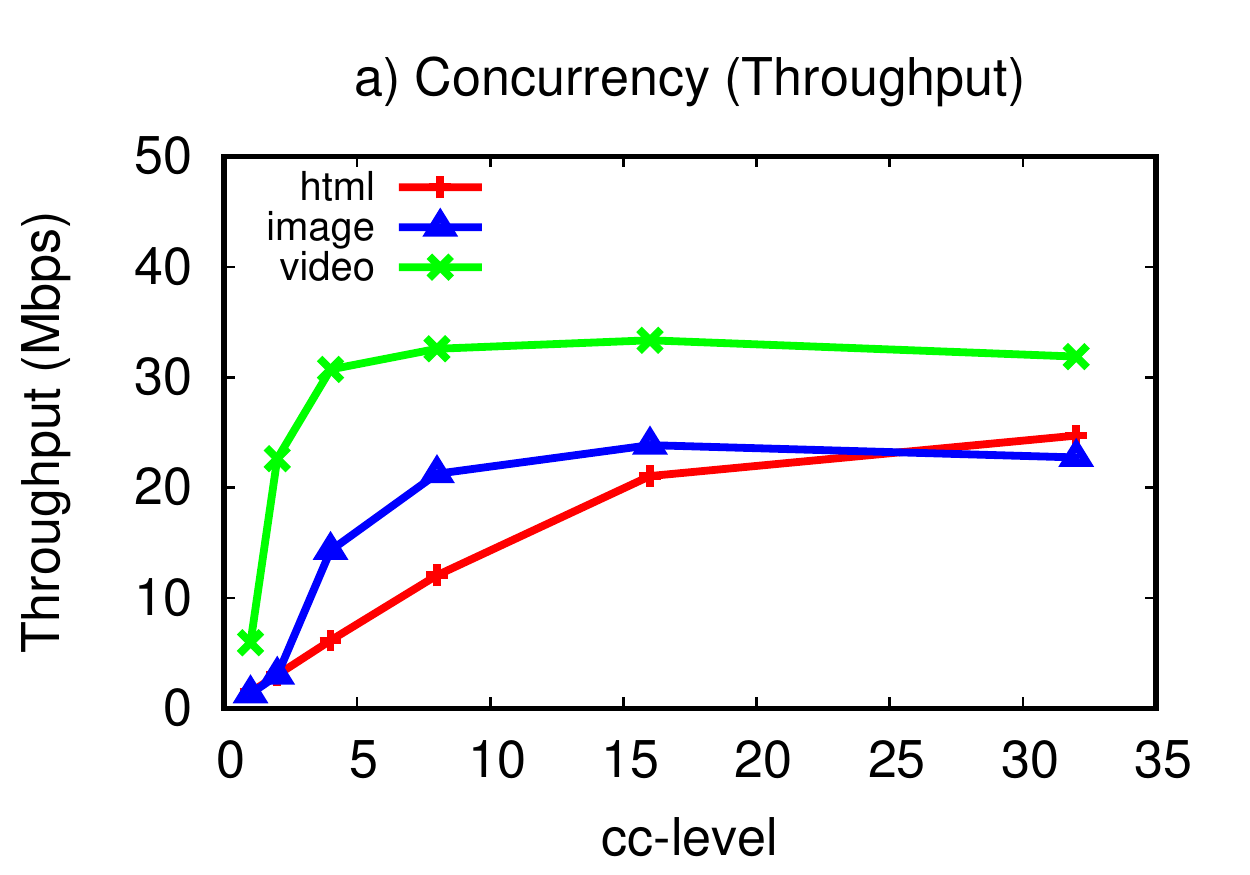}
						\hspace{1cm}
			\includegraphics[keepaspectratio=true,angle=0,width=65mm]{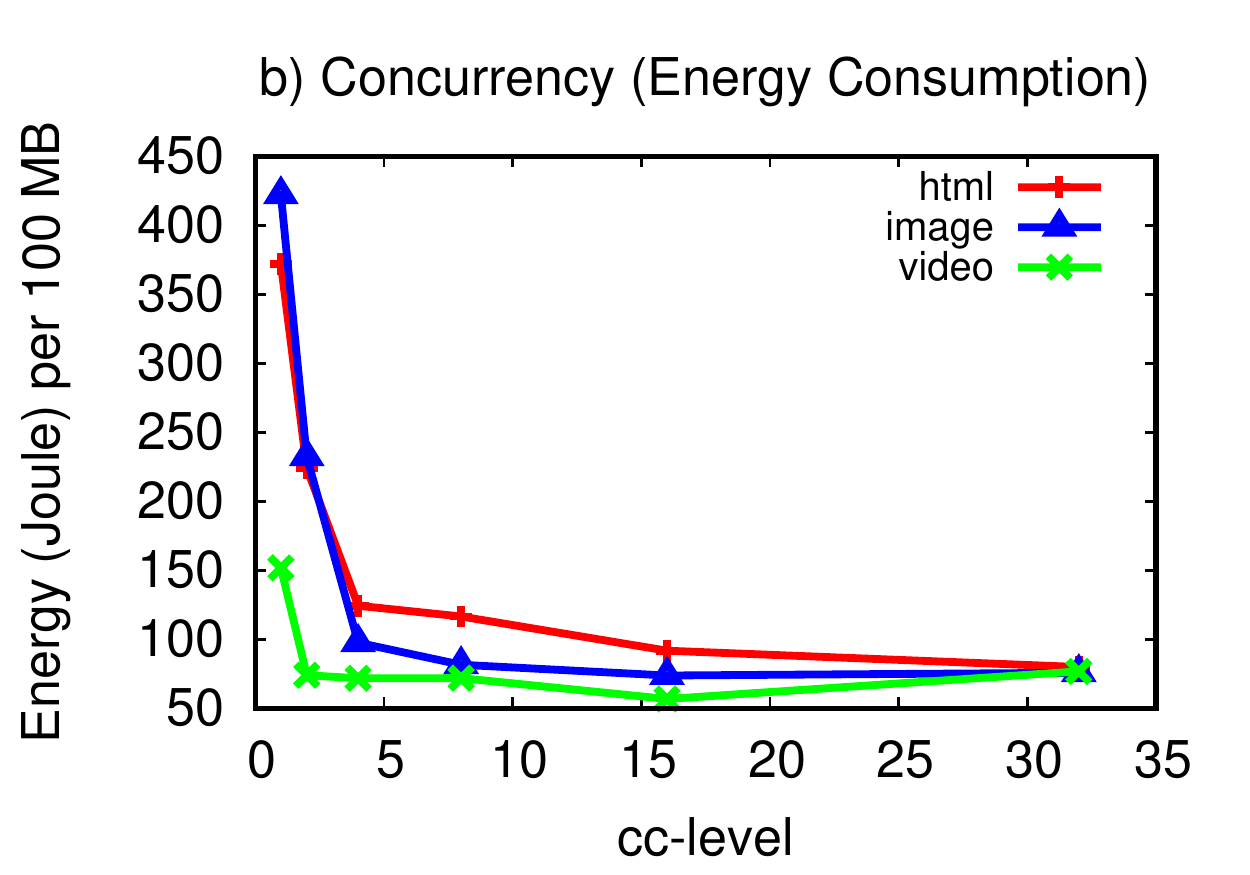}\\
			\includegraphics[keepaspectratio=true,angle=0,width=65mm]{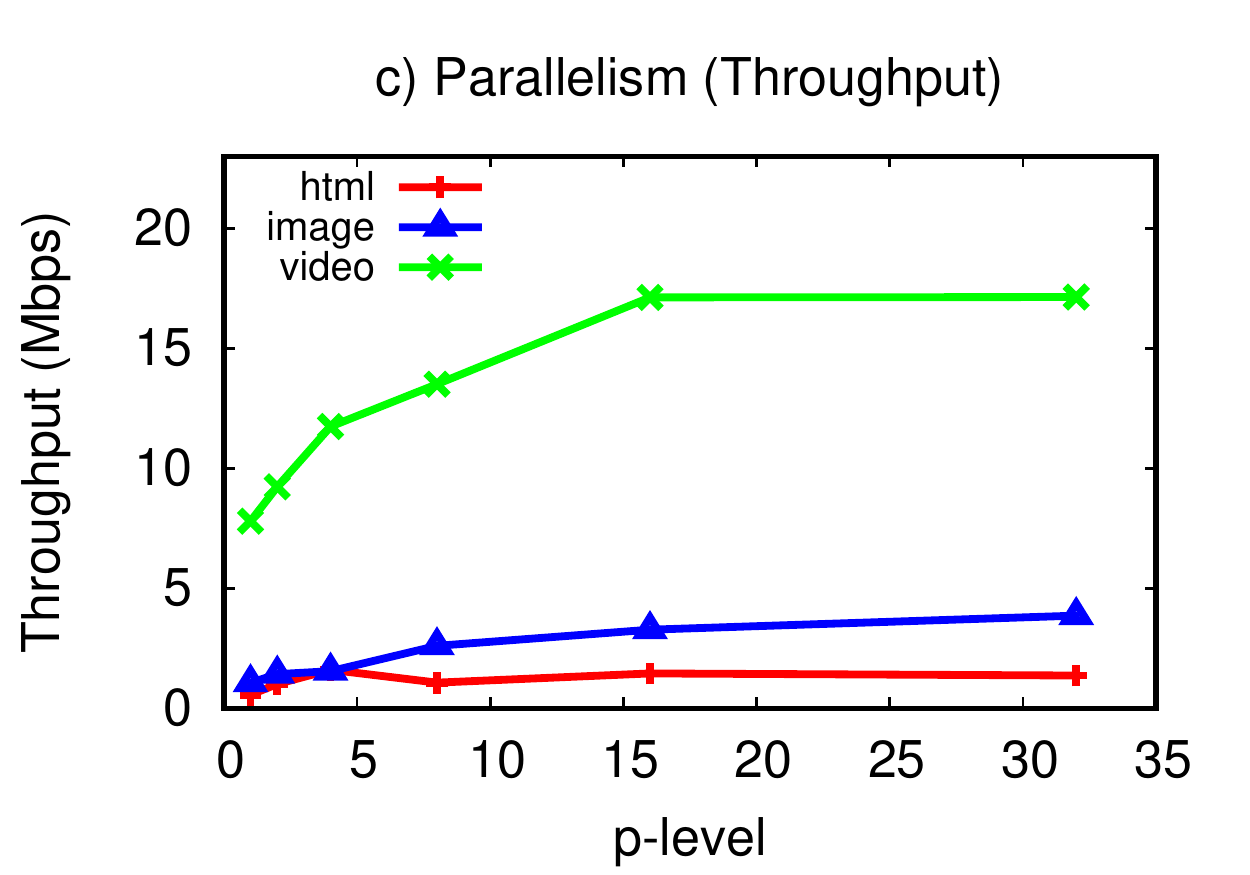}  
						\hspace{1cm}
			\includegraphics[keepaspectratio=true,angle=0,width=65mm]{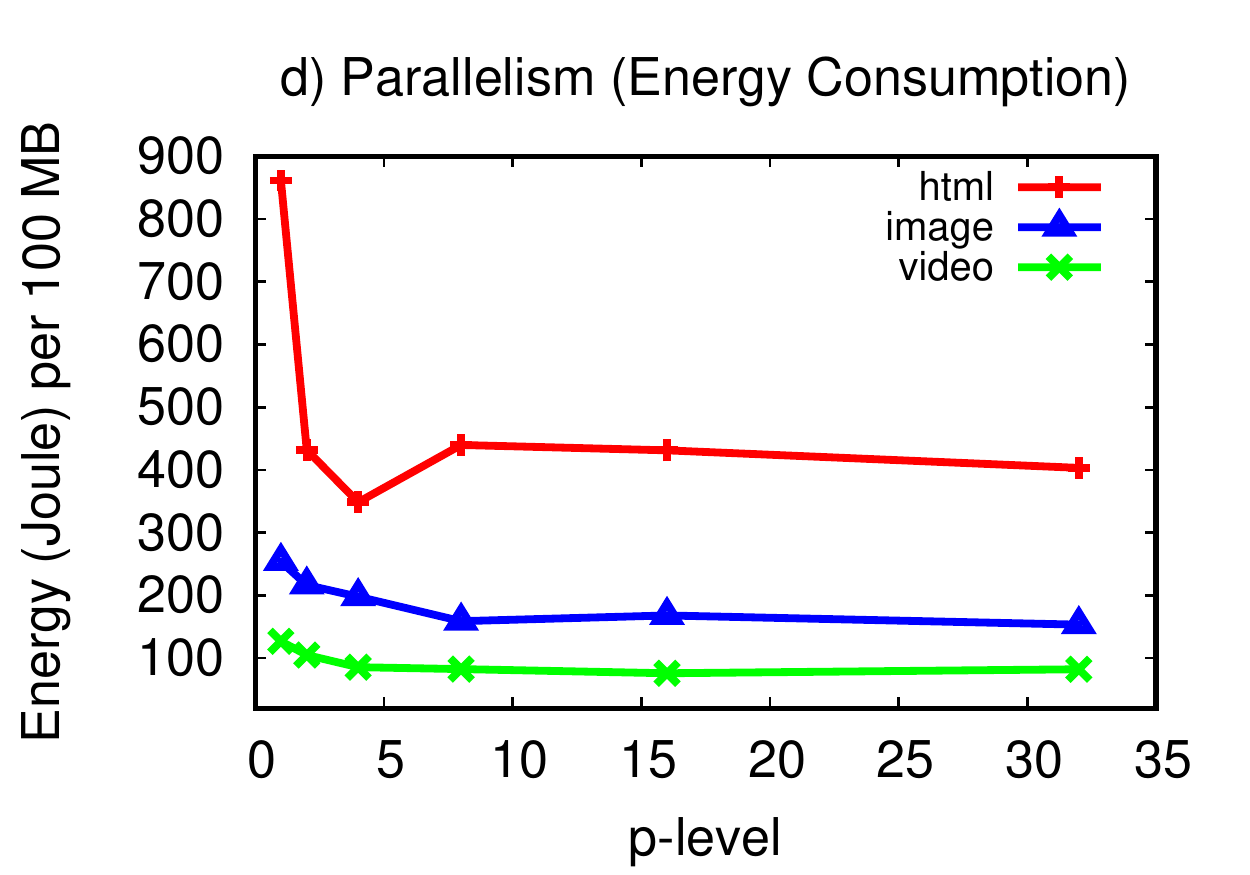}
		\end{tabular}
		\caption{Throughput vs Energy Consumption trade-offs of individual protocol parameters for 4G/LTE data transfers between AWS EC2 Sydney and DIDCLAB Galaxy S5.} \label{fig:4Gparameter}
	\end{centering}
	\vspace{-5mm}
\end{figure*}

Overall, concurrency parameter showed a better performance than parallelism on our html, image and video datasets. When we increased level of concurrency from 1 to 32, it boosted end-to-end throughput for html, image and video datasets and reduced energy consumption on the smartphone client as seen in Figure~\ref{fig:parameter} (a)-(b). As we increase concurrency level from 1 to 16 for html dataset, the throughput almost doubled at each level and increased from 2.6 Mbps to 42 Mbps, which is 16.7X improvement. For the image dataset, the throughput increases from 5.6 Mbps to 48 Mbps (8.5X improvement), and for the video dataset it increases from 13 Mbps to 61 Mbps (4.5X improvement). Increasing the level of concurrency from 1 to 16 reduced total energy consumption 70\% for the html dataset, 68\% for the image dataset, and 38\% for the video dataset. After concurrency level 16, it still continued to keep energy saving balanced for html dataset while it started to increase the energy consumption for image and video Datasets. On the other hand, when it comes to the parallelism parameter, the performance of each dataset showed different characteristics. Increased level of parallelism improved the end-to-end throughput of the video dataset transfers and decreased the energy consumption up to a specific level as shown at Figure~\ref{fig:parameter} (c)-(d). As the parallelism level increased, the throughput improved 3.1X and energy consumption decreased 45\%. On the other hand, parallelism did not improve the throughput for html and image datasets. In fact, up to parallelism level 8, it gradually reduced the end-to-end throughput of html and image datasets and increased the energy consumption. After this level, increasing parallelism slightly improved throughput for both dataset, but it still continued to increase energy consumption for the html dataset while it is reduced for the image dataset. 

We examined the reasons that could cause the negative effect of parallelism on the html and image datasets. It is known that parallelism can be more beneficial when the buffer size is smaller than the Bandwidth-Delay-Product (BDP), which especially occurs in large bandwidth and long RTT networks~\cite{YildirimPCP}. We run extra data transfers for analyzing the effect of parallelism on larger files. Figure~\ref{fig:th-pw10G} shows the further analysis of individual file size effect of parallelism on throughput and energy consumption per 100 MB transfers from AWS EC2 Sydney web server and the client Samsung Galaxy S5 at DIDCLAB. We used three specific file sizes for this test: 1 GB, 3GB, and 10 GB. As we increased the level of parallelism from 1 to 32, the throughput of each transfer improved and energy consumption decreased. The throughput of 1 GB, 3 GB, 10 GB files increased from 24 Mbps to 86 Mbps (3.5X improvement), 24 Mbps to 92 Mbps (3.75X improvement), and 24 Mbps to 104 Mbps (4.4X improvement) respectively. Energy consumption rates of 1 GB, 3 GB, 10 GB files per 100 MB transfers also decreased 32\%, 33\%, and 35\% respectively up to parallelism level 16. These results show that in order to take advantage of parallelism, bigger individual file sizes should be adopted by considering buffer size of the system and the BDP.

\begin{figure*}[t]
	\begin{centering}
		\hspace*{-0.2cm}\begin{tabular}{ccc}
			\includegraphics[keepaspectratio=true,angle=0,width=58mm]{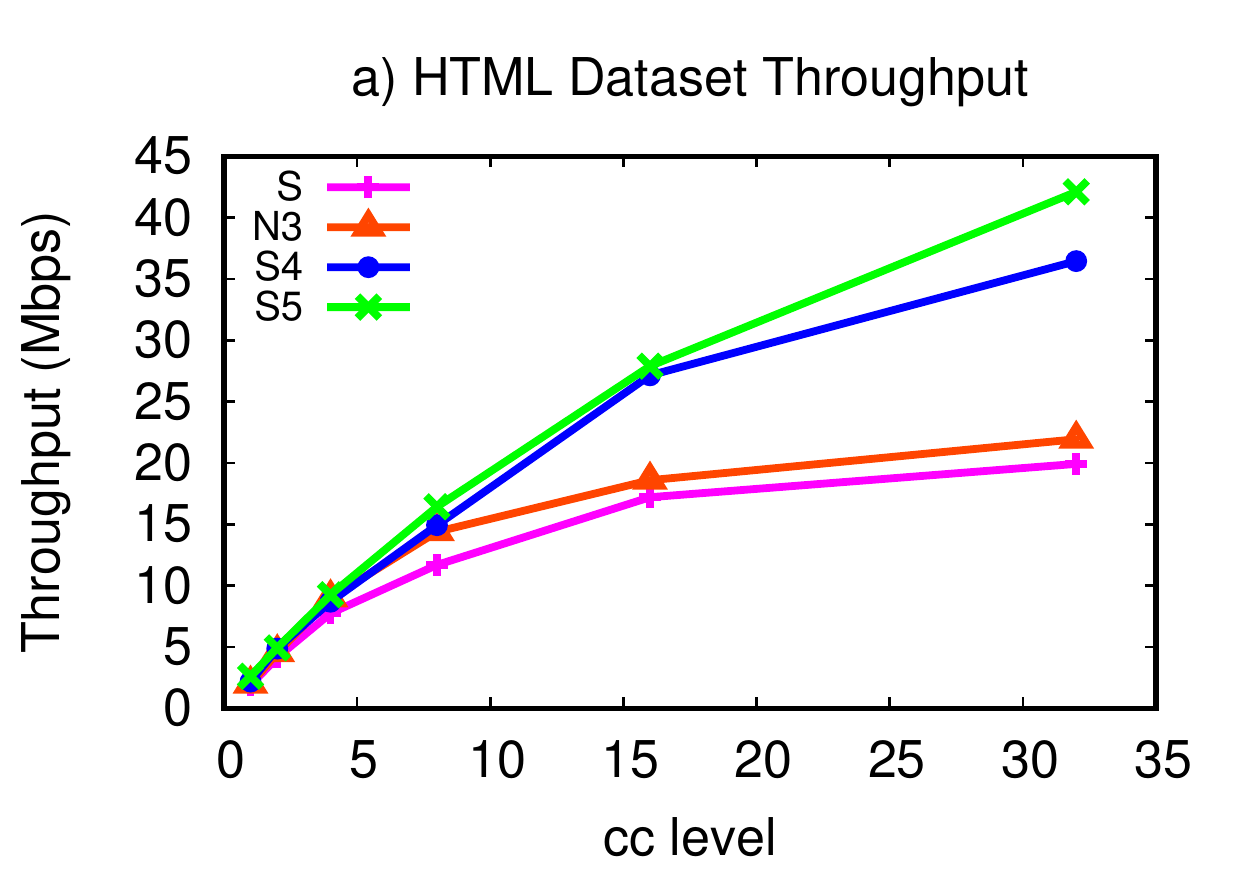}
			\includegraphics[keepaspectratio=true,angle=0,width=58mm]{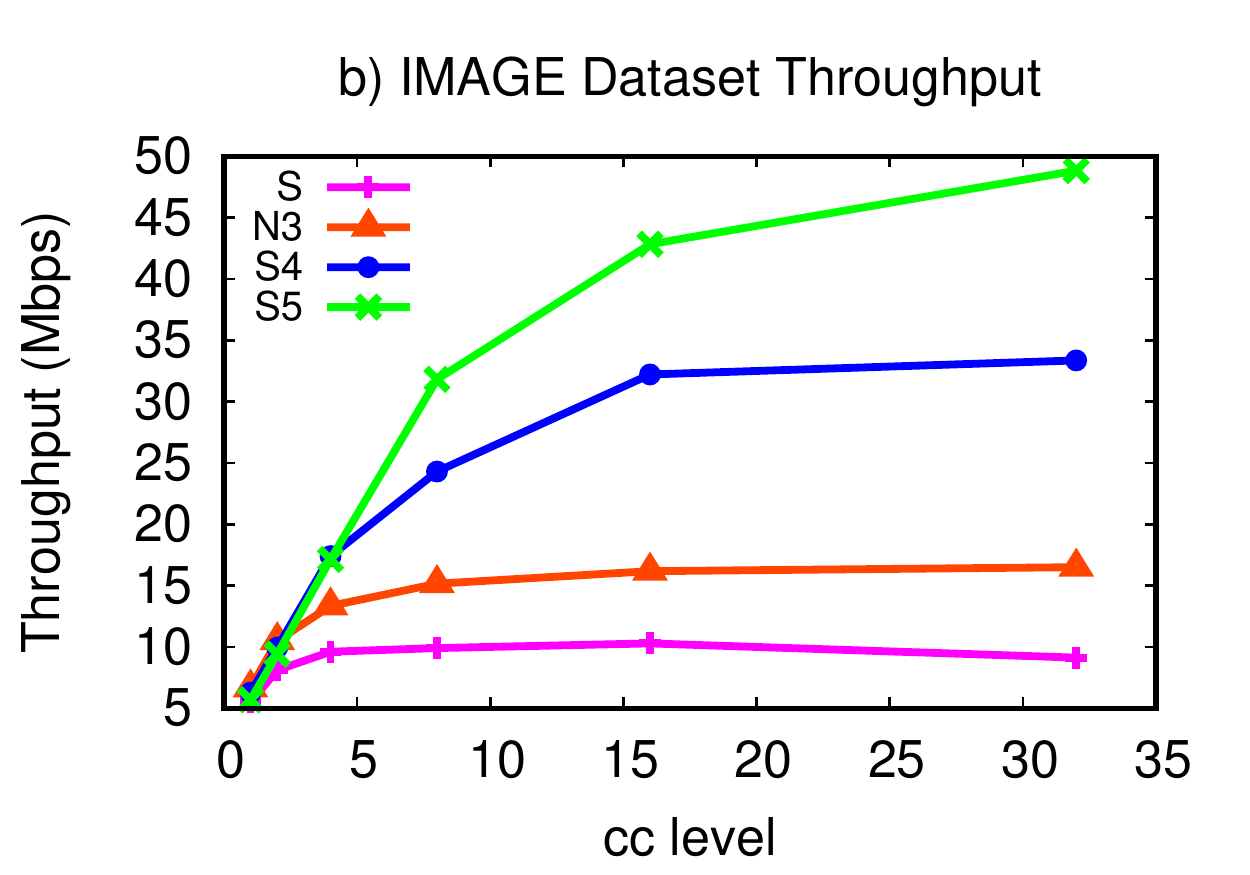}
			\includegraphics[keepaspectratio=true,angle=0,width=58mm]{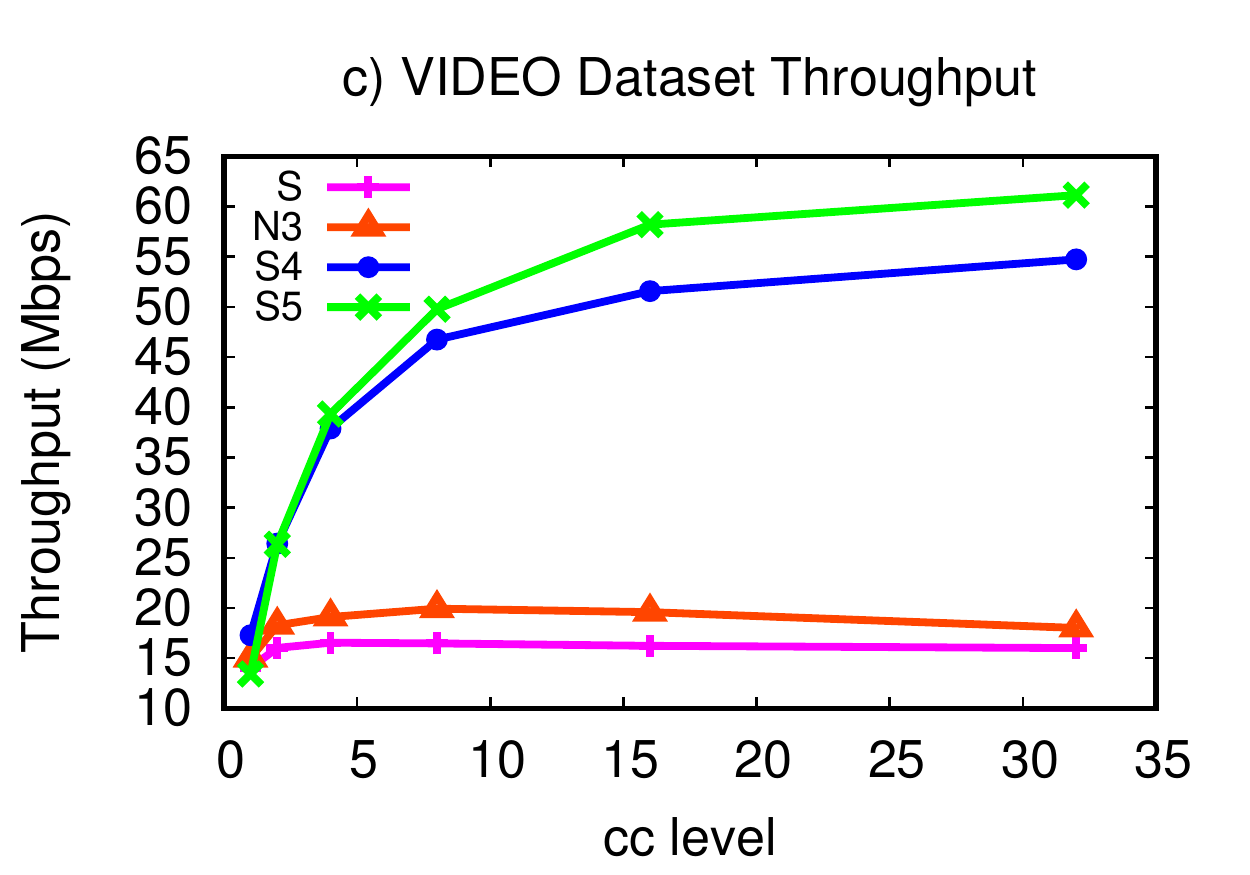}
			\\ 
			\includegraphics[keepaspectratio=true,angle=0,width=58mm]{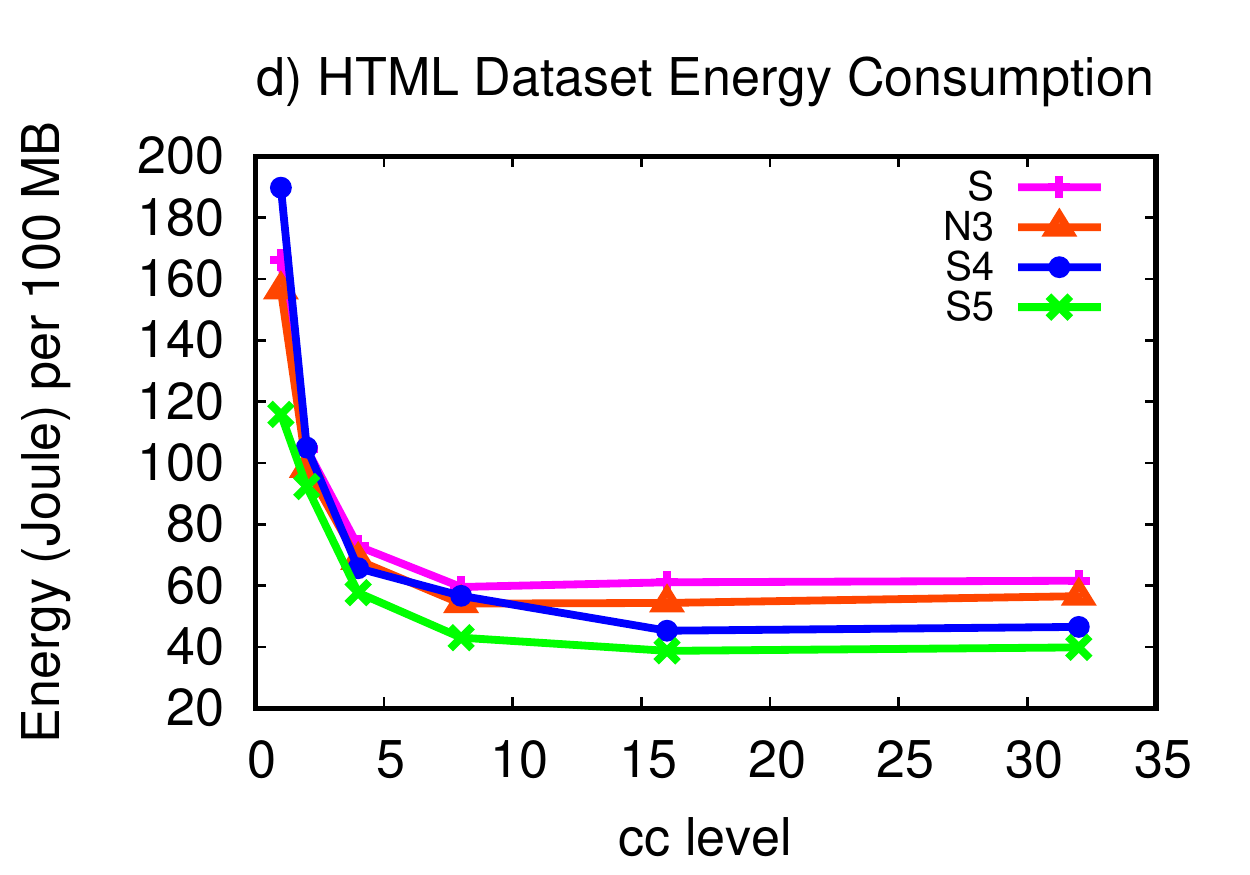} 
			\includegraphics[keepaspectratio=true,angle=0,width=58mm]{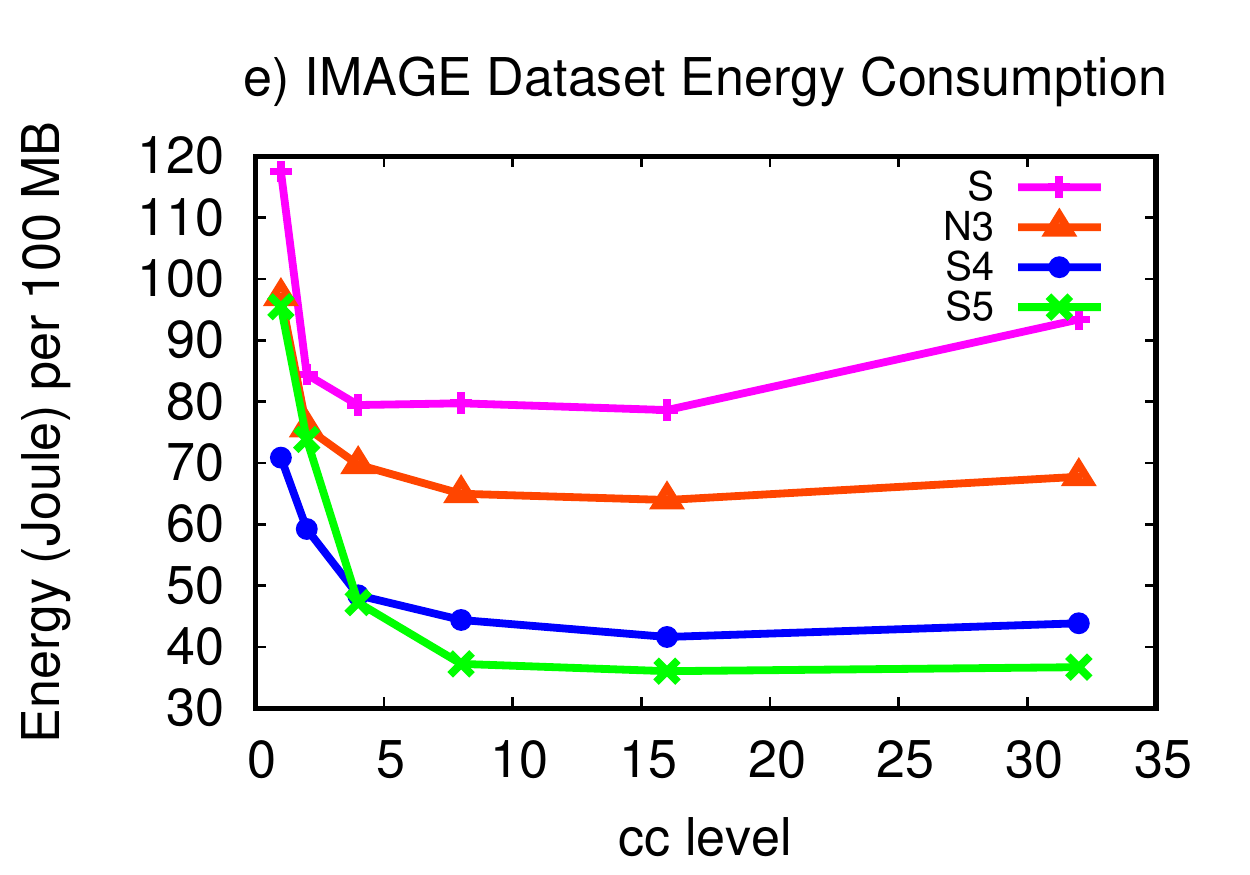}
			\includegraphics[keepaspectratio=true,angle=0,width=58mm]{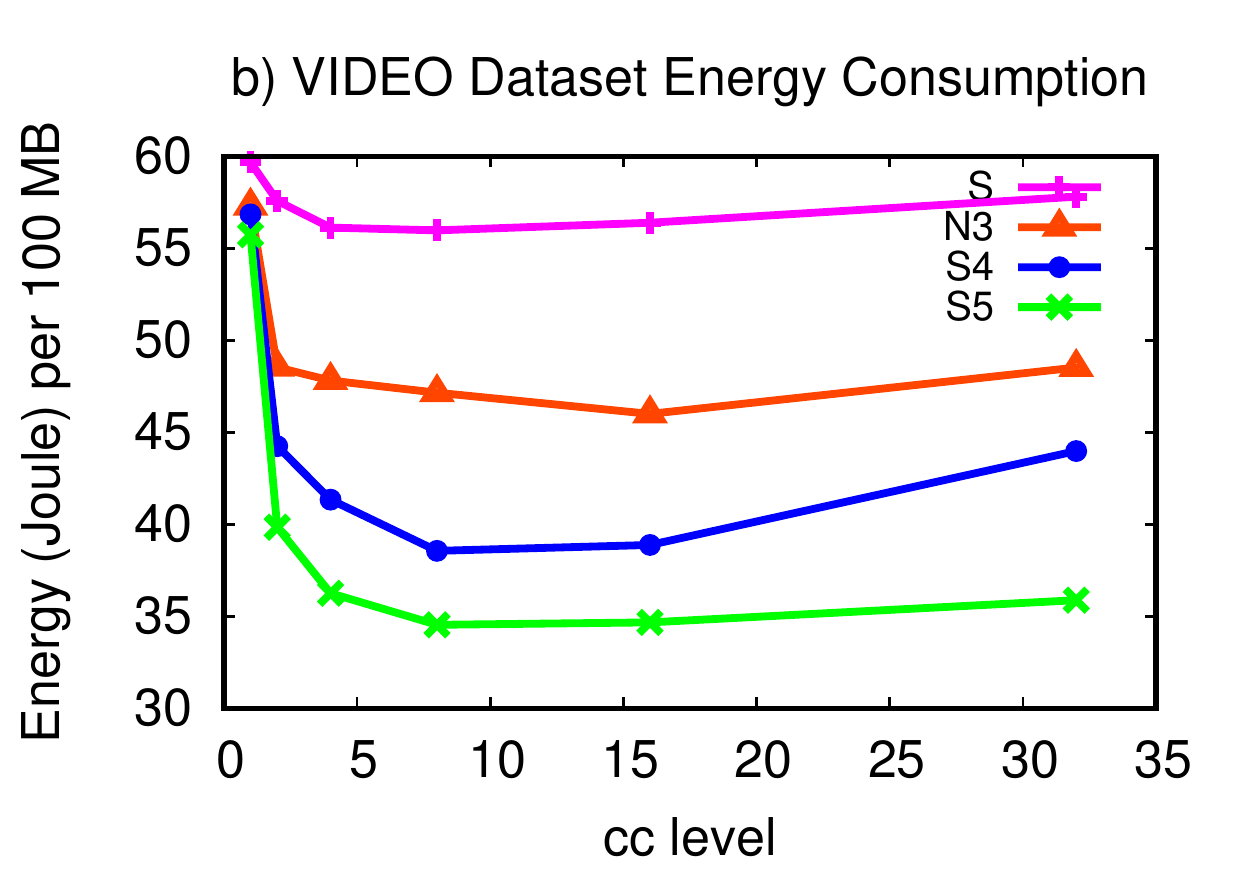}
\end{tabular}
	\caption{Throughput vs Energy Consumption per 100 MB data transfer  from AWS EC2 Sydney server to different mobile devices at DIDCLAB.} 		\label{fig:phones}
	\end{centering}
\end{figure*}

\begin{figure*}[!htb]
    \begin{centering}
		\hspace*{-0.2cm}\begin{tabular}{ccc}
			\includegraphics[keepaspectratio=true,angle=0,width=58mm]{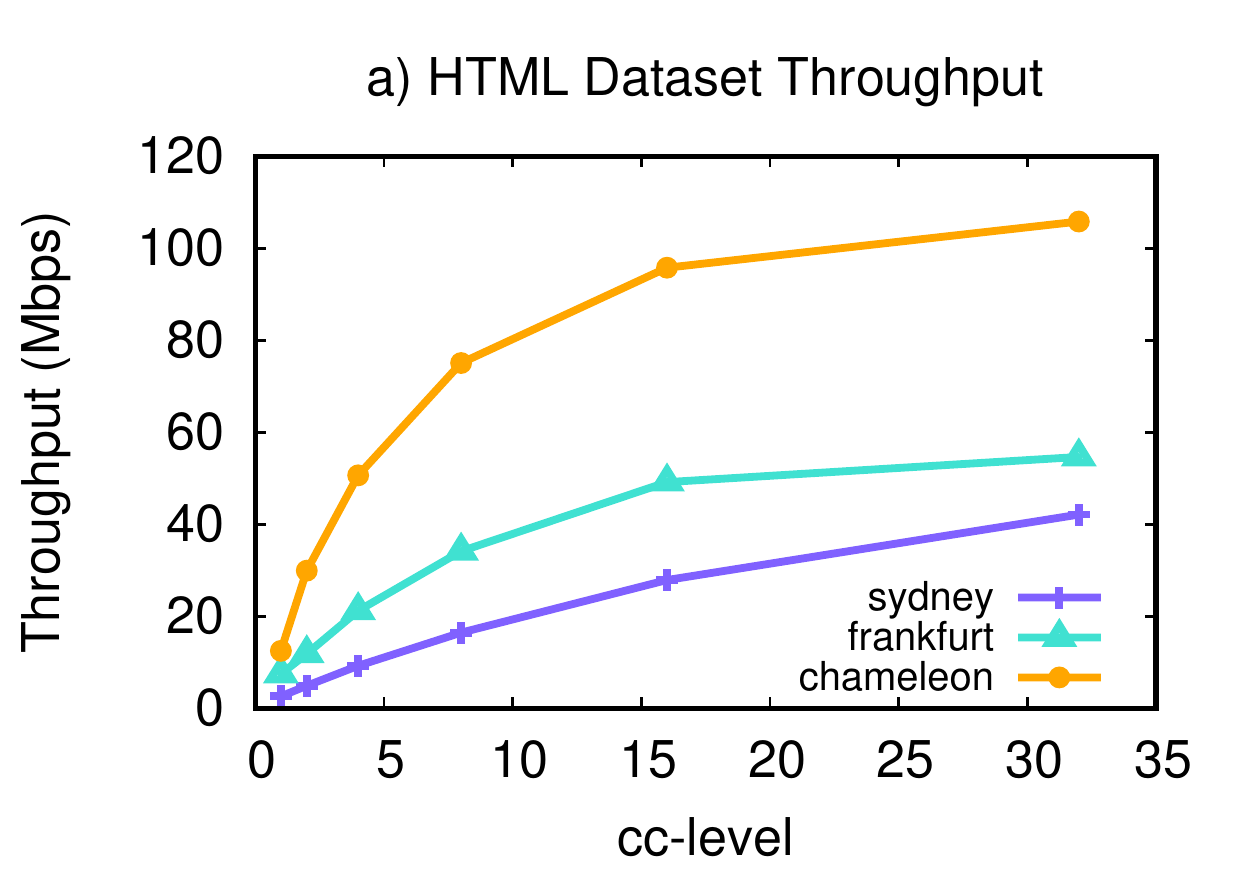}
			\includegraphics[keepaspectratio=true,angle=0,width=58mm]{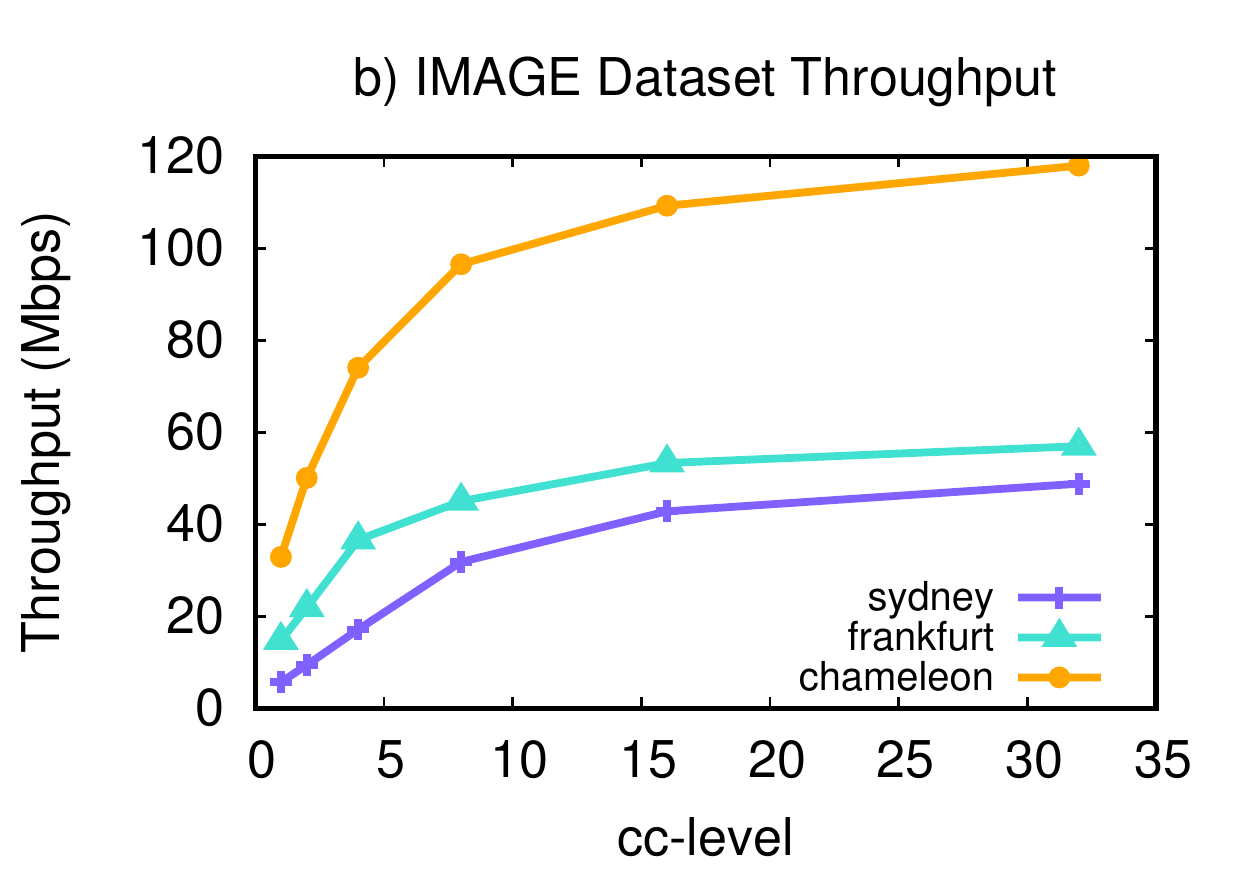}
			\includegraphics[keepaspectratio=true,angle=0,width=58mm]{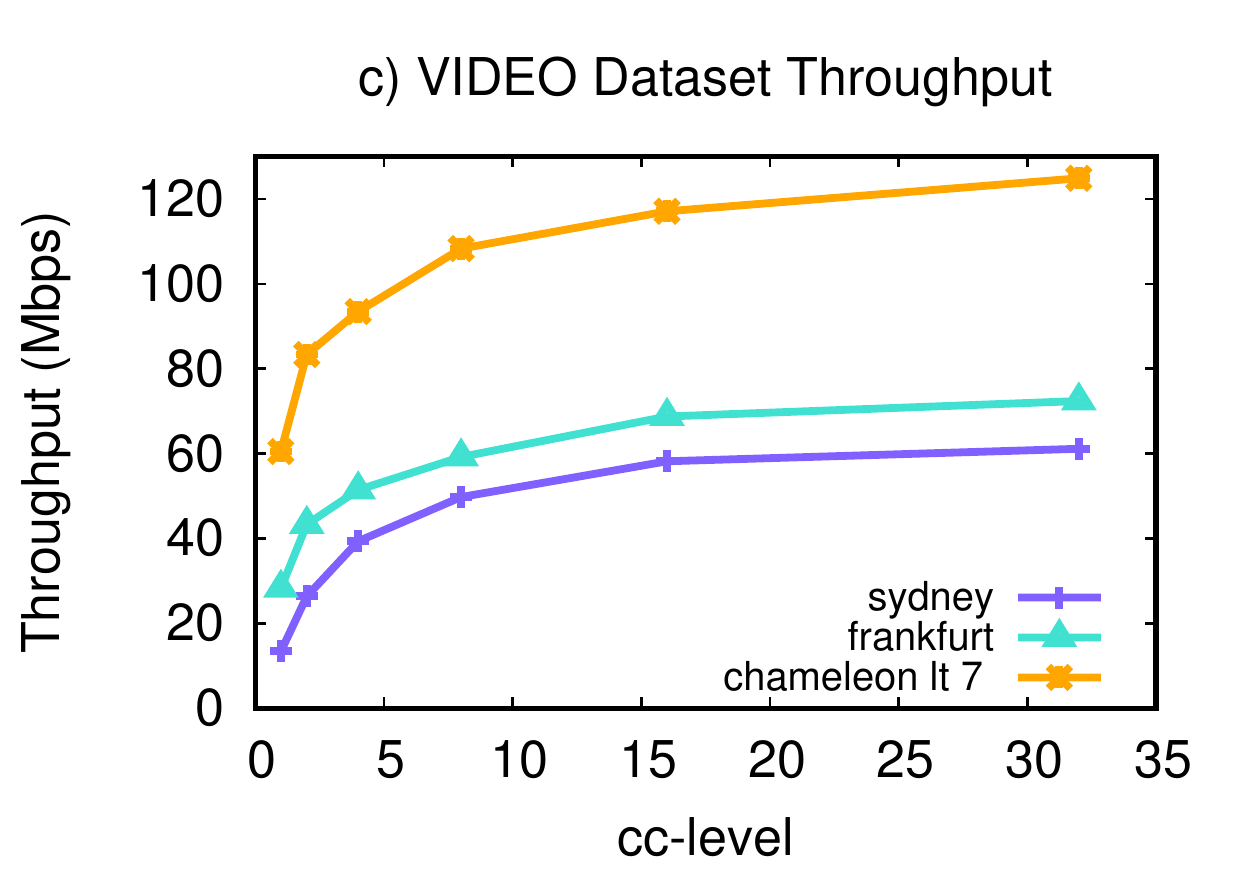}
			\\ 
			\includegraphics[keepaspectratio=true,angle=0,width=58mm]{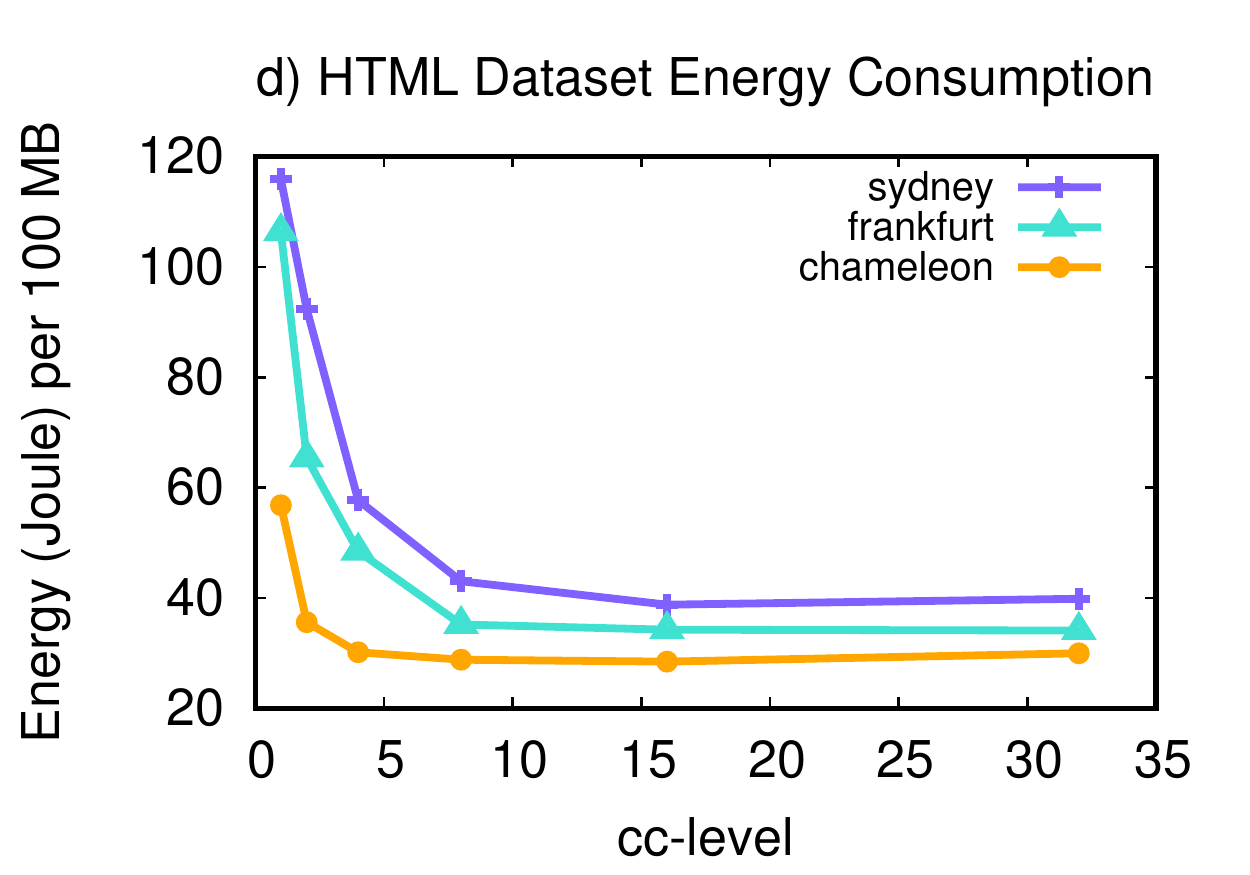} 
			\includegraphics[keepaspectratio=true,angle=0,width=58mm]{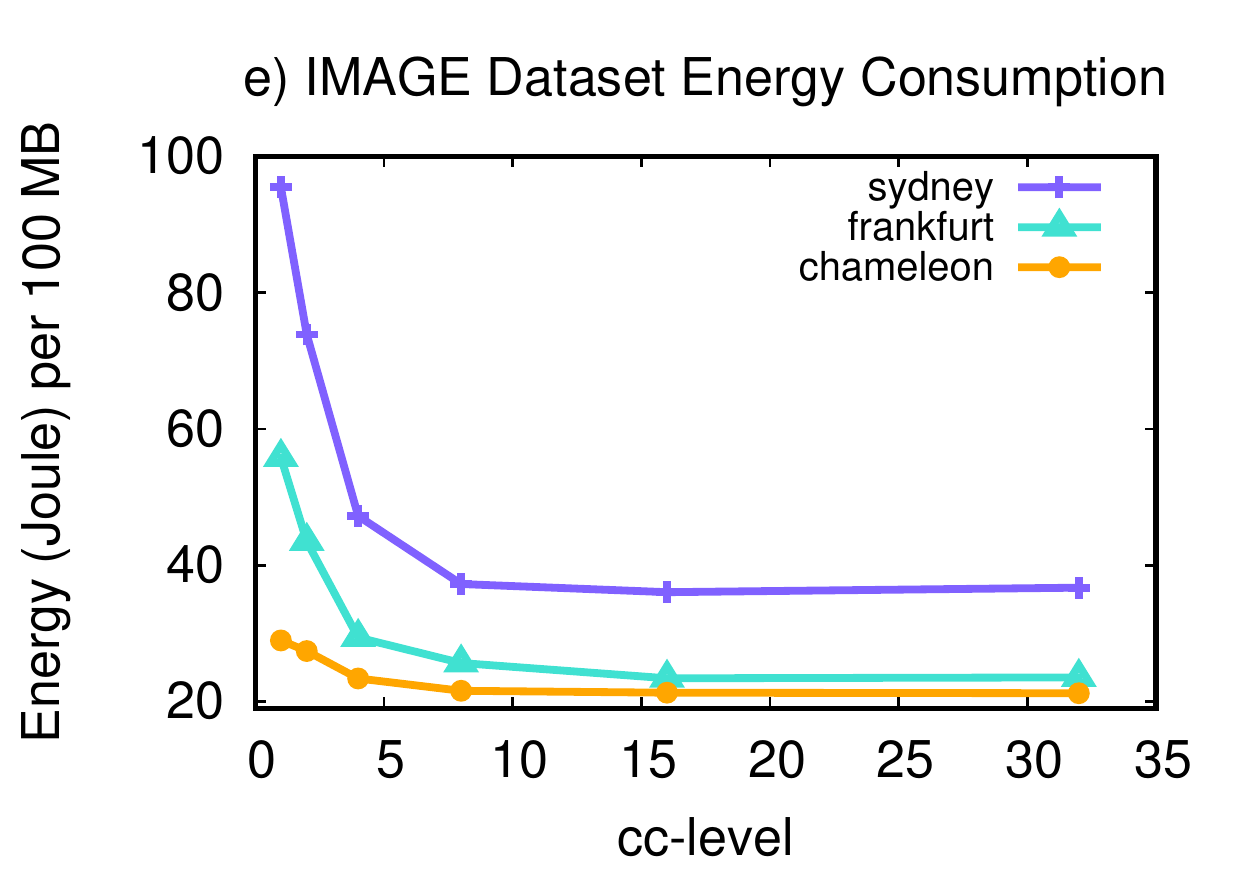}
			\includegraphics[keepaspectratio=true,angle=0,width=58mm]{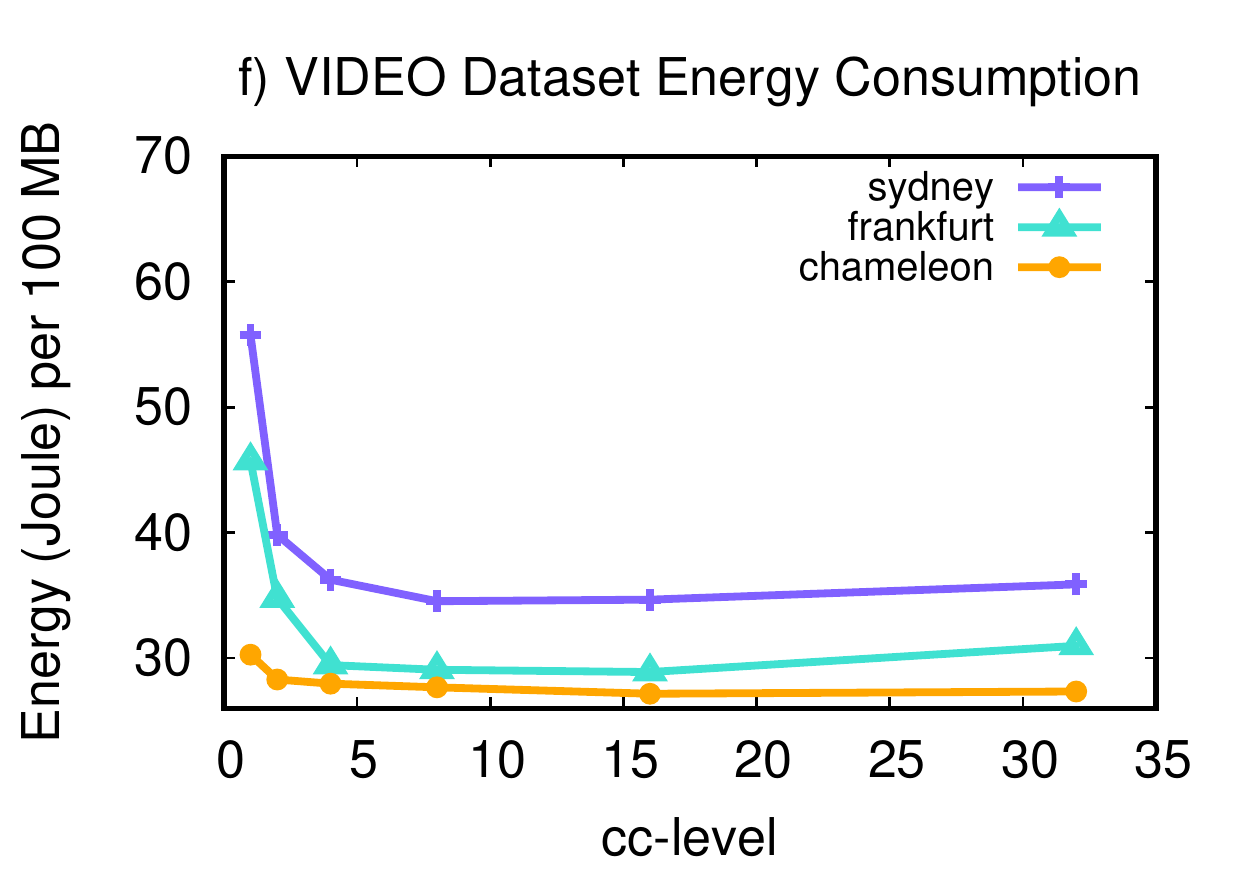}
		\end{tabular}
        \caption{Throughput vs Energy Consumption per 100 MB data transfer  from different servers to Samsung Galaxy S5 at DIDLAB.} \label{fig:servers}
	\end{centering}
\end{figure*}

There are other ways of improving end-to-end data transfer throughput and saving energy other than tuning network parameters (i.e. concurrency and parallelism) without interfering end-systems. Choosing optimal I/O request size of the application is a good example of one of these ways as seen in Figure~\ref{fig:parameter} (e)-(f). With high speed networks, end-systems may become the bottleneck in terms of responding this links, and sometimes a small tune up in application's writing to disk speed makes a noticeable change in overall performance. By doing a smart I/O request size tune up on smartphones that have limited memory and storage space, similarly can increase overall end-to-end throughput. 

\begin{figure*}[!htb]
    \begin{centering}
		\begin{tabular}{cc}
			\includegraphics[keepaspectratio=true,angle=0,width=50mm]{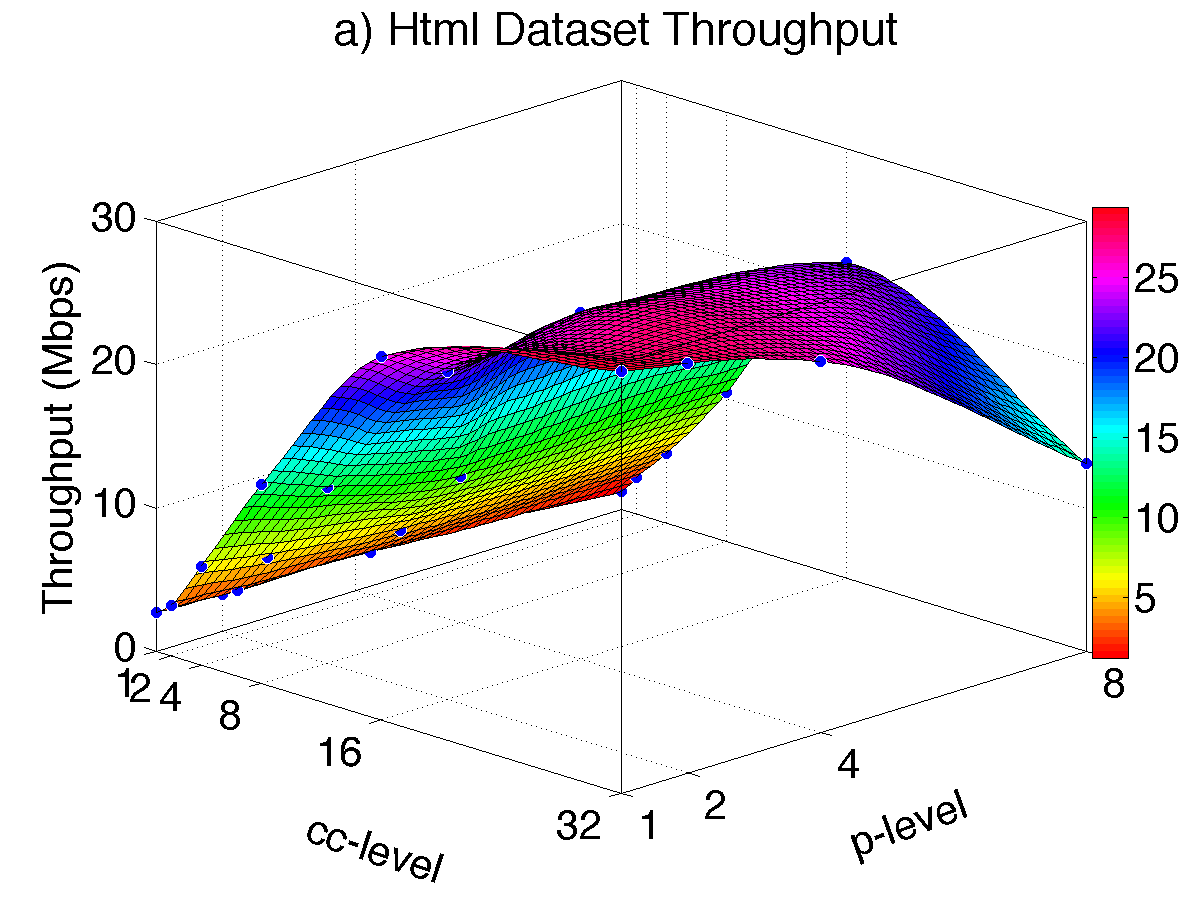}
			\vspace{0.5cm}
			\hspace{2cm}
			\includegraphics[keepaspectratio=true,angle=0,width=50mm]{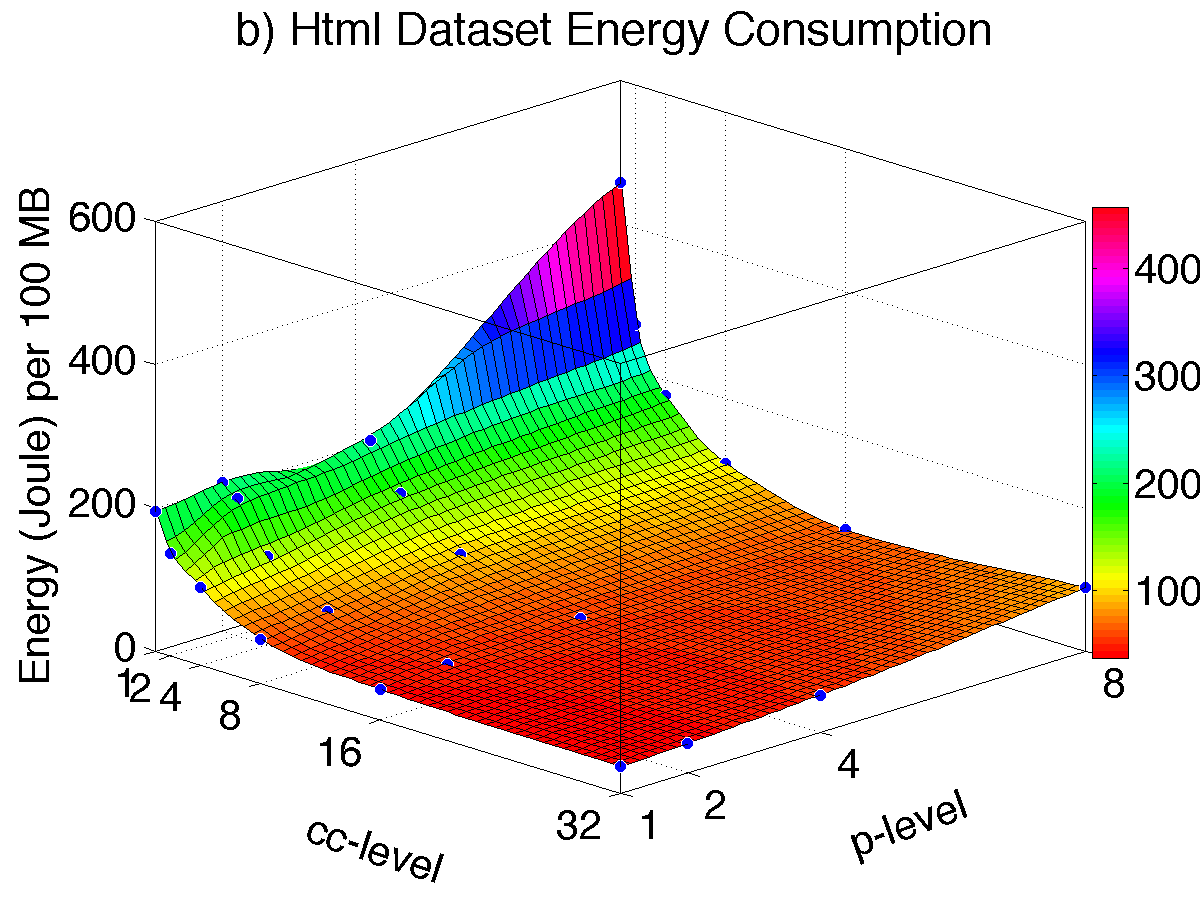}\\
			\includegraphics[keepaspectratio=true,angle=0,width=50mm]{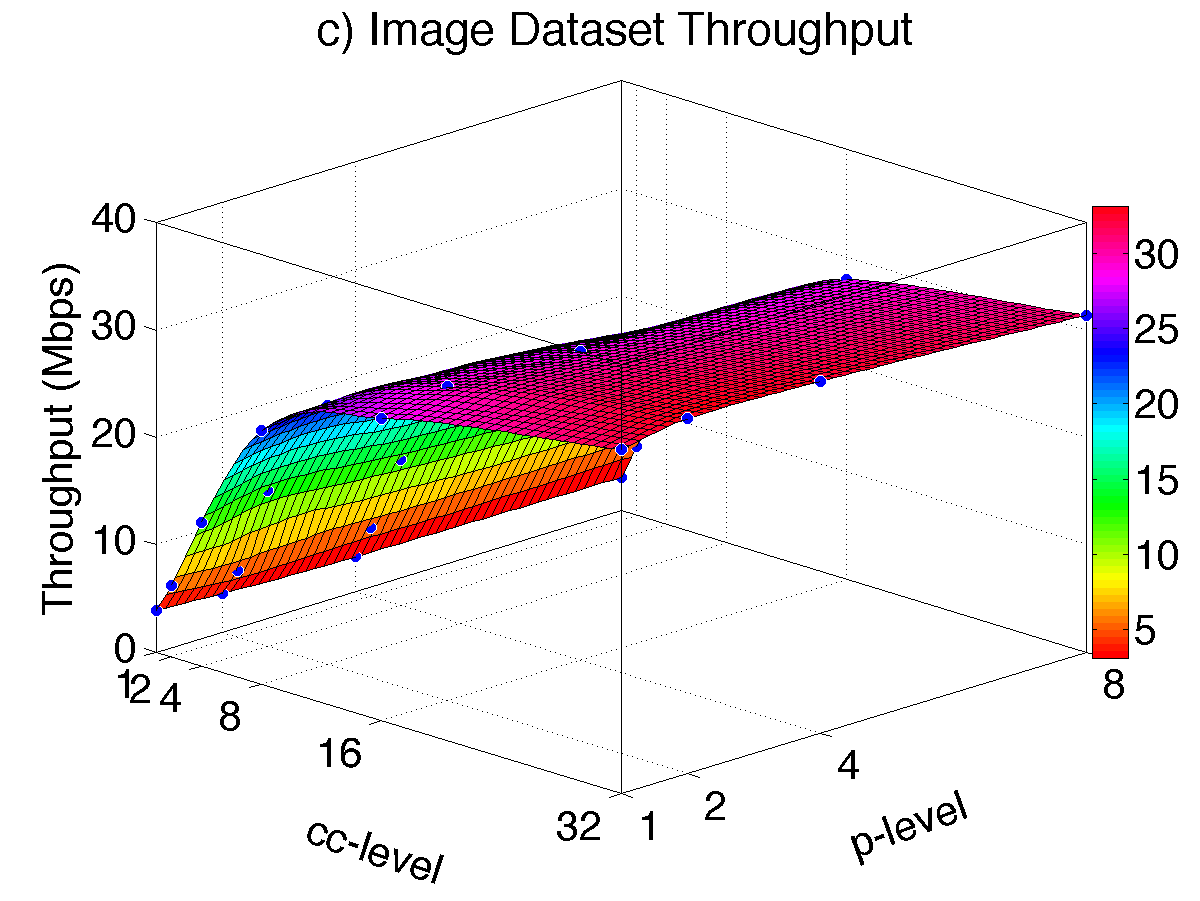}
			\vspace{0.5cm}
			\hspace{2cm}
			\includegraphics[keepaspectratio=true,angle=0,width=50mm]{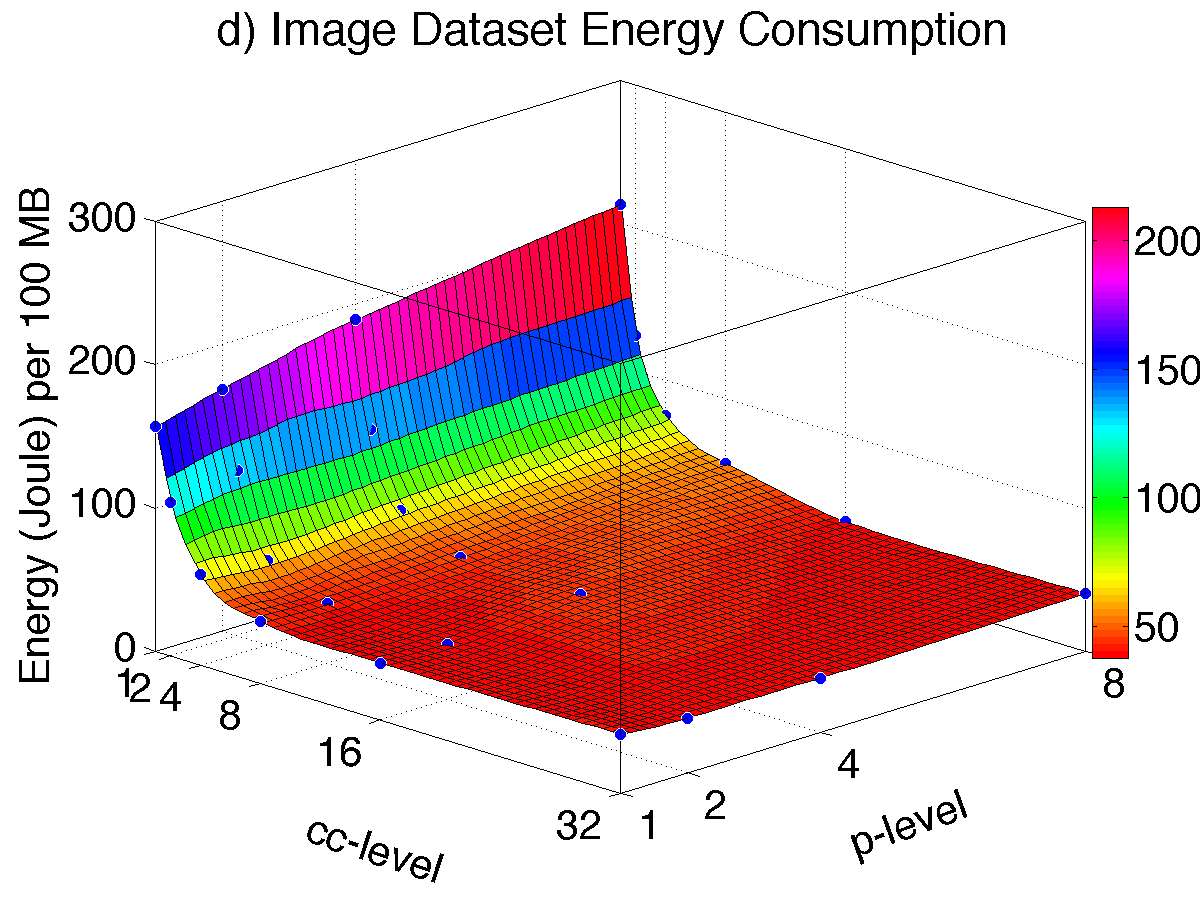} \\
			\includegraphics[keepaspectratio=true,angle=0,width=50mm]{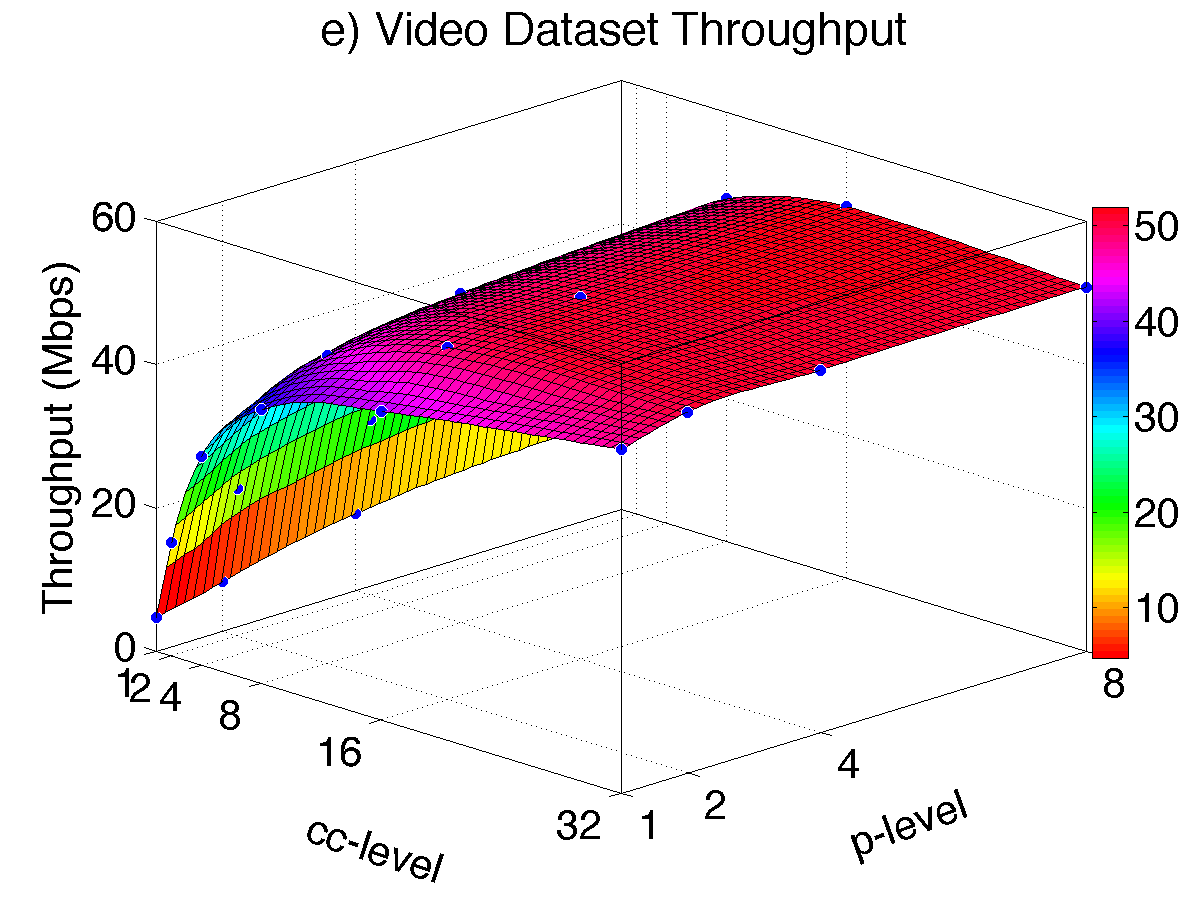}
			\vspace{0.5cm} 
			\hspace{2cm}
			\includegraphics[keepaspectratio=true,angle=0,width=50mm]{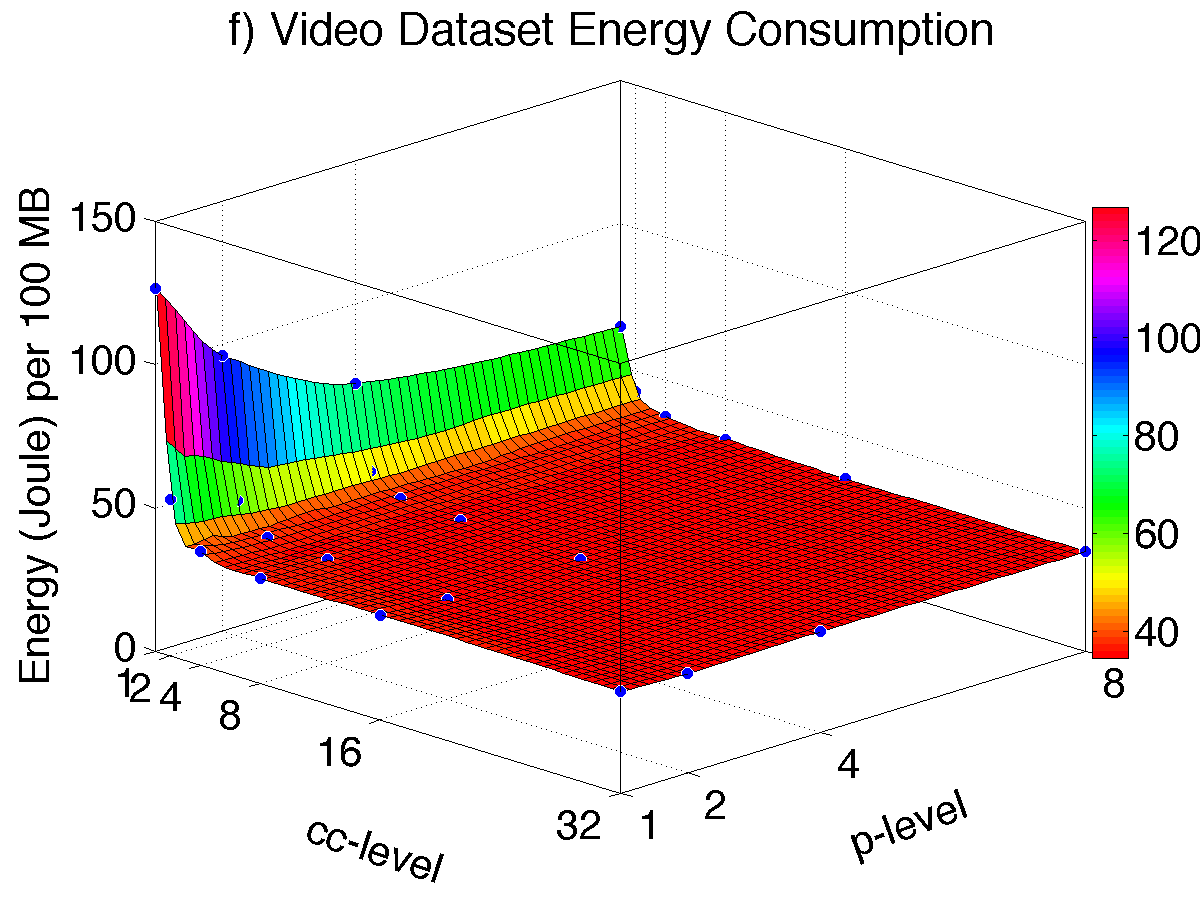}\\
			\includegraphics[keepaspectratio=true,angle=0,width=50mm]{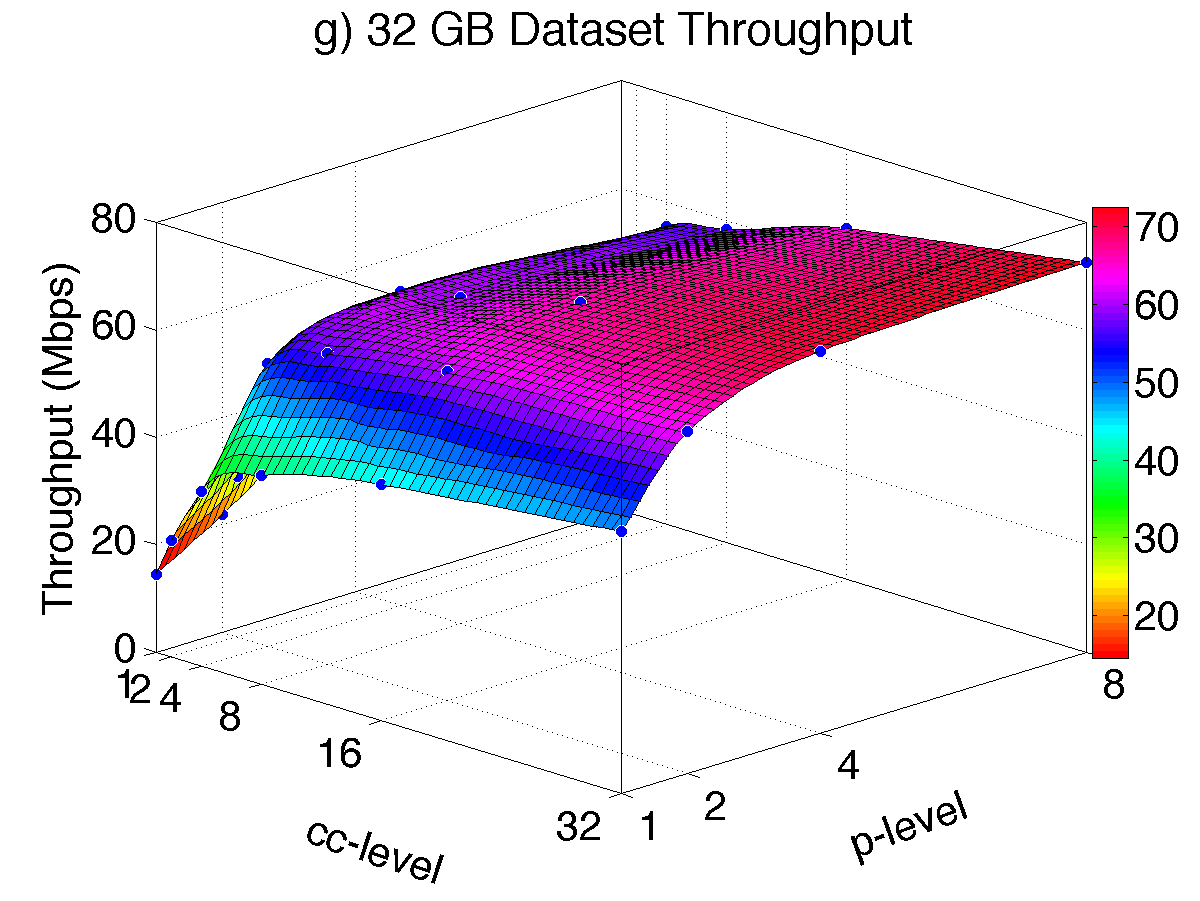} 
			\vspace{0.5cm}
			\hspace{2cm}
			\includegraphics[keepaspectratio=true,angle=0,width=50mm]{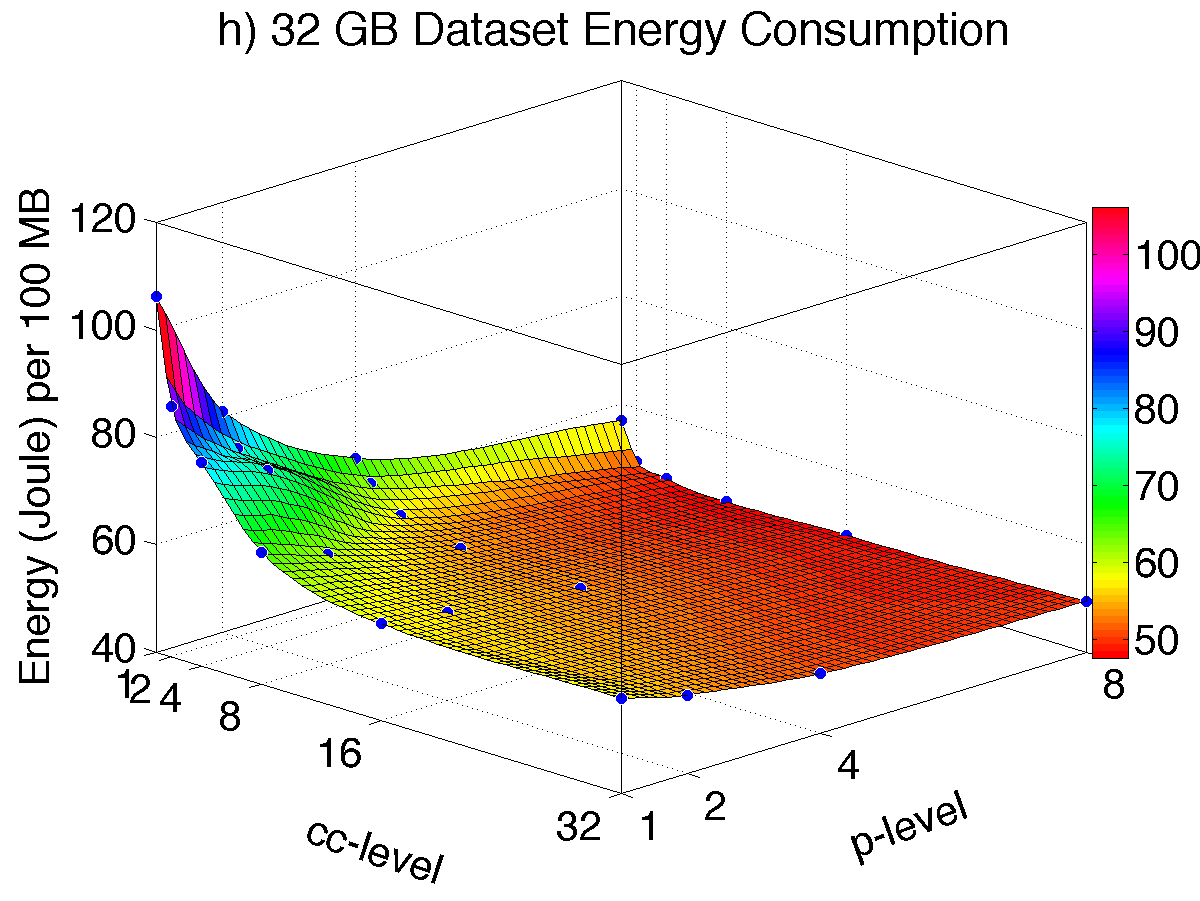}
		\end{tabular}
			\vspace{-1mm}
        \caption{Throughput vs Energy Consumption trade-offs of combined protocol parameters for WiFi data transfers from AWS EC2 Sydney to DIDCLAB Galaxy S5.} \label{fig:matlabS5}
	\end{centering}
\end{figure*}

After reaching optimal I/O request size of the application, further increase either does not change the already balanced system or causes a slight decrease. To give a better understanding, we set up multiple experiments to test our smartphone application's I/O request size effect on throughput and energy consumption during the html, image and video dataset transfers with concurrency and parallelism parameters. Figure~\ref{fig:parameter} (e)-(f) shows the change in throughput and energy consumption during video dataset transfers from AWS EC2 Sydney server to the client Galaxy S5 at DIDCLAB. We doubled I/O block size from 1 KB to 64 KB for all experiments. 
When I/O request size is increased from 1 KB to 8-16 KB, throughput slightly increased and came to a balance. Further increasing I/O request size induced increase in energy consumption at same concurrency level, besides, it did not increase throughput. The main reason for this is that the main bottleneck during the end-to-end data transfers in our testbed was not the mobile client's storage I/O speed, rather it was the wireless (WiFi or 4G LT) network connectivity.

\begin{figure*}[t]
	\begin{centering}
		\begin{tabular}{cc}
			\includegraphics[keepaspectratio=true,angle=0,width=65mm]{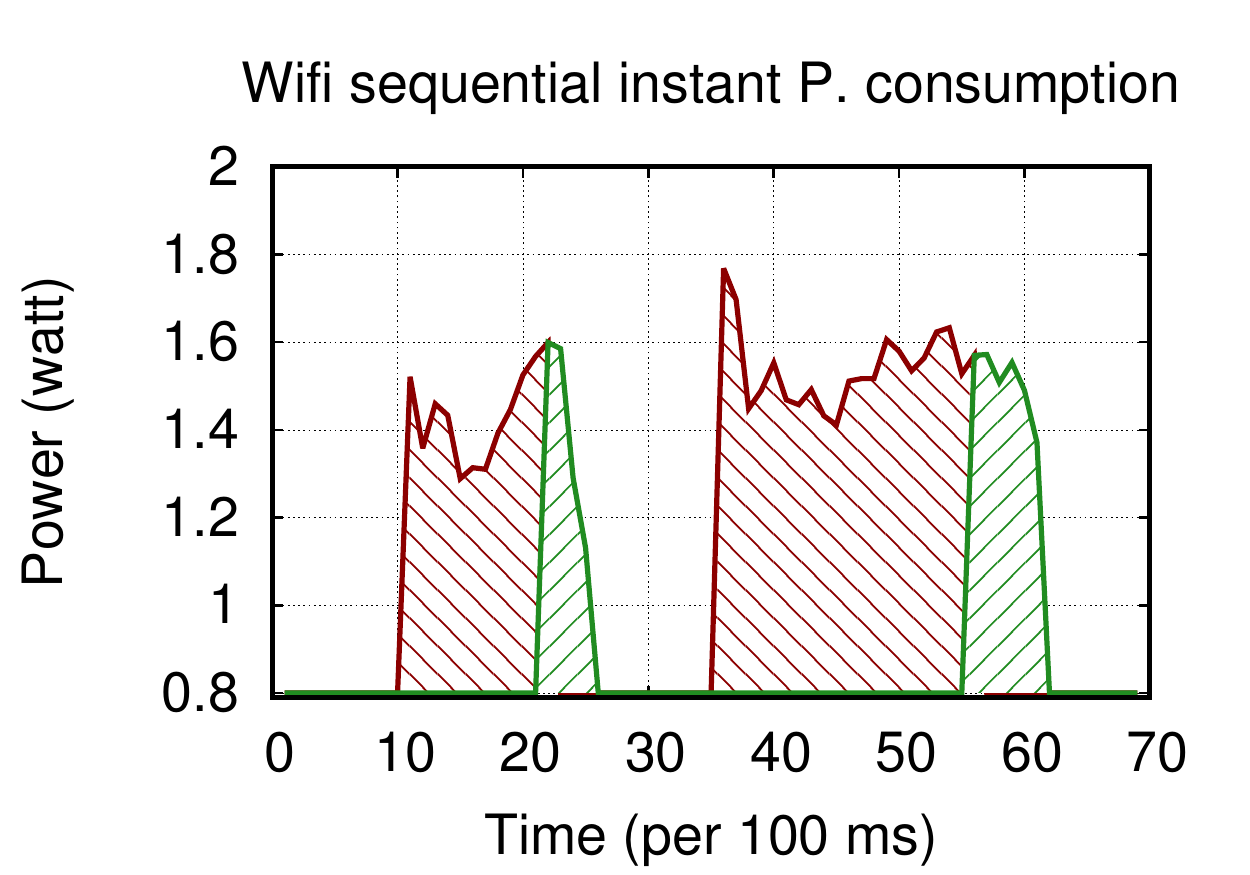}
			\hspace{1cm}
			\includegraphics[keepaspectratio=true,angle=0,width=65mm]{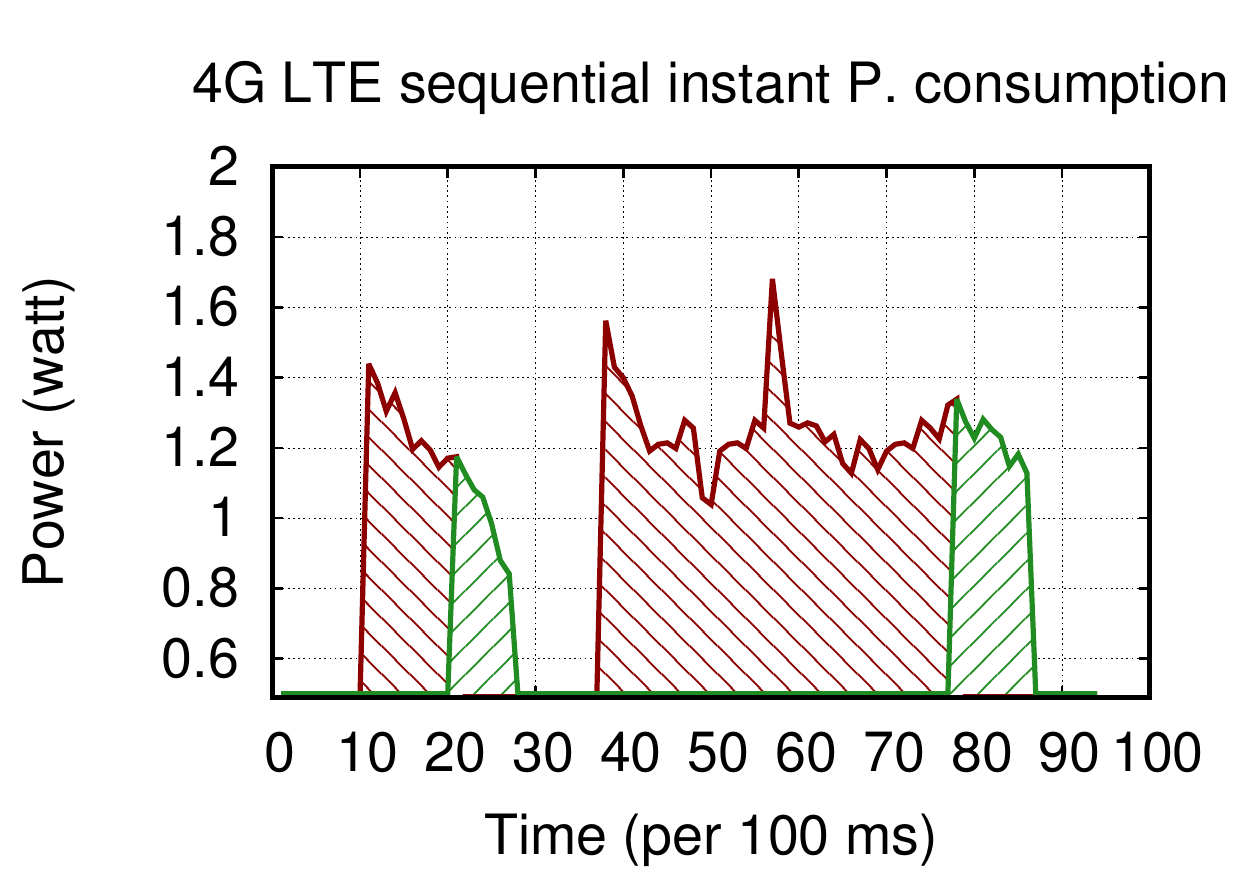}\\        
			\includegraphics[keepaspectratio=true,angle=0,width=65mm]{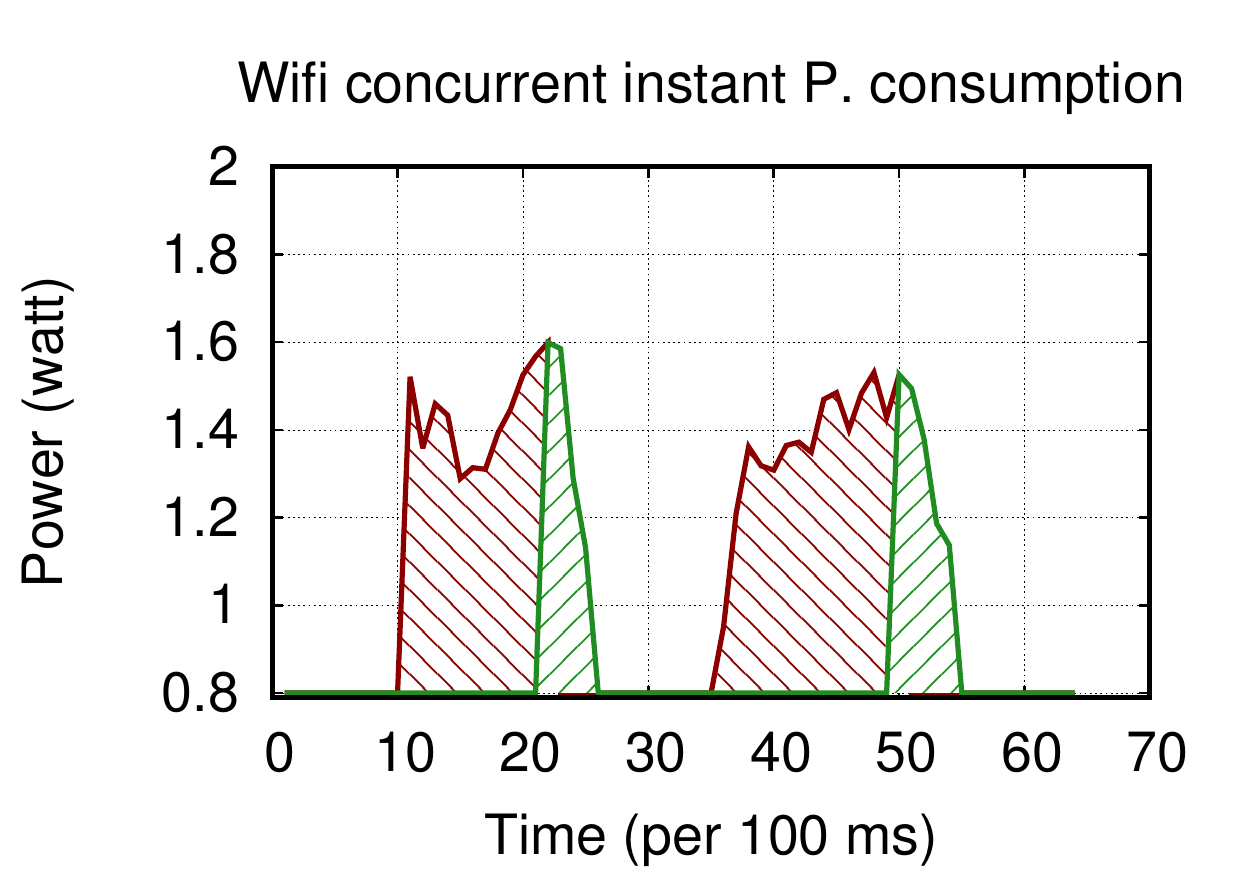}
			\hspace{1cm}
			\includegraphics[keepaspectratio=true,angle=0,width=65mm]{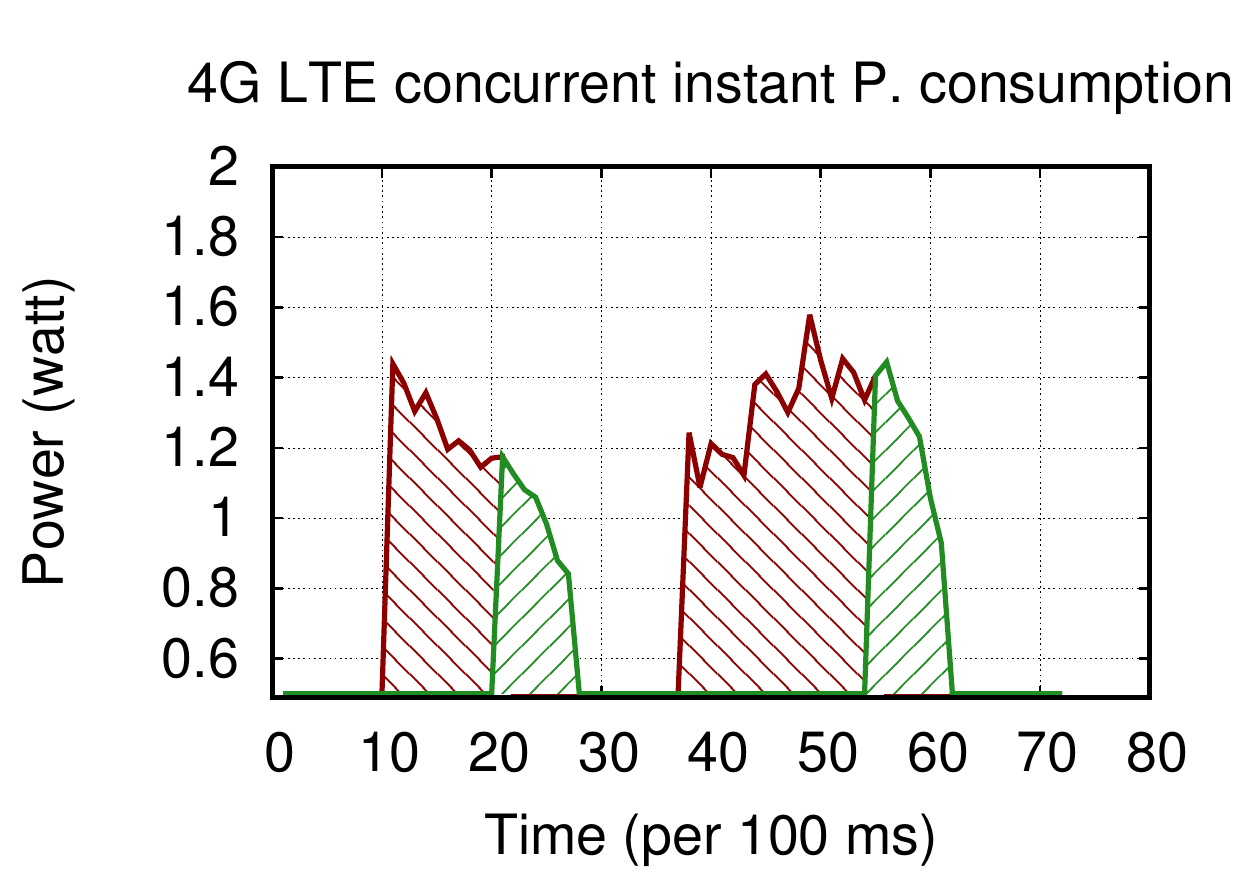}
		\end{tabular}
		\caption{Instantaneous power consumption graph showing actual data transfer energy (red) versus tail energy (green) during sequential and concurrent single and double file transfers with WiFi and 4G LTE.} 
		\label{fig:instant4Gwifi}
	\end{centering}
\end{figure*}

Having throughput and energy consumption results of individual protocol parameters with WiFi connection, we also run the same individual parameter experiments with 4G LTE as presented in Figure~\ref{fig:4Gparameter}. Considering download/upload speeds of the cellular networks and WiFi, the results were quite similar, but with slightly lower end-to-end data transfer throughput. The noticeable difference was in the effect of parallelism on energy saving. Comparing the image and html dataset results with WiFi connection, increased level of parallelism had positively effected the energy saving with 4G LTE on these two datasets. This is mainly caused by the speed of the 4G LTE network. While the speed of 4G LTE network did not outdo achieved maximum throughput performance of concurrency parameter of WiFi connection, it increased the energy efficiency of html, image and video datasets. Comparing with WiFi connection, the energy saving increased from 70\% to 77\% for html, from 68\% to 76\% for image, and  from 38\% to 39.5\% for the video dataset. Again, increased level of concurrency became more effective for all datasets. Both parallelism and concurrency improved the end-to-end data transfer throughput and energy saving for all three datasets on 4G LTE data transfers.

\begin{figure*}[t]
	\begin{centering}
	\begin{tabular}{cc}
			\vspace{-5mm}
			\includegraphics[keepaspectratio=true,angle=0,width=55mm]{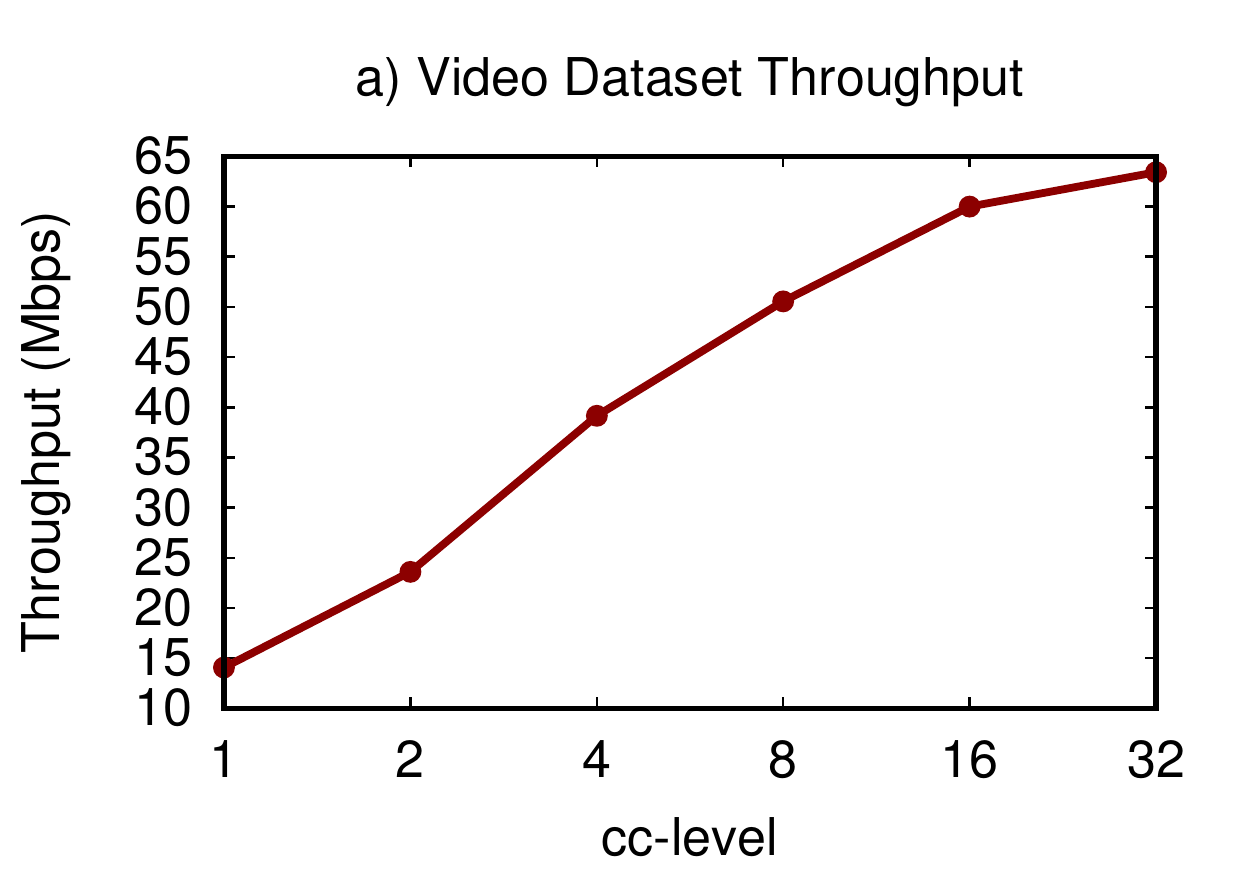}
			\hspace{1cm}
			\includegraphics[keepaspectratio=true,angle=0,width=55mm]{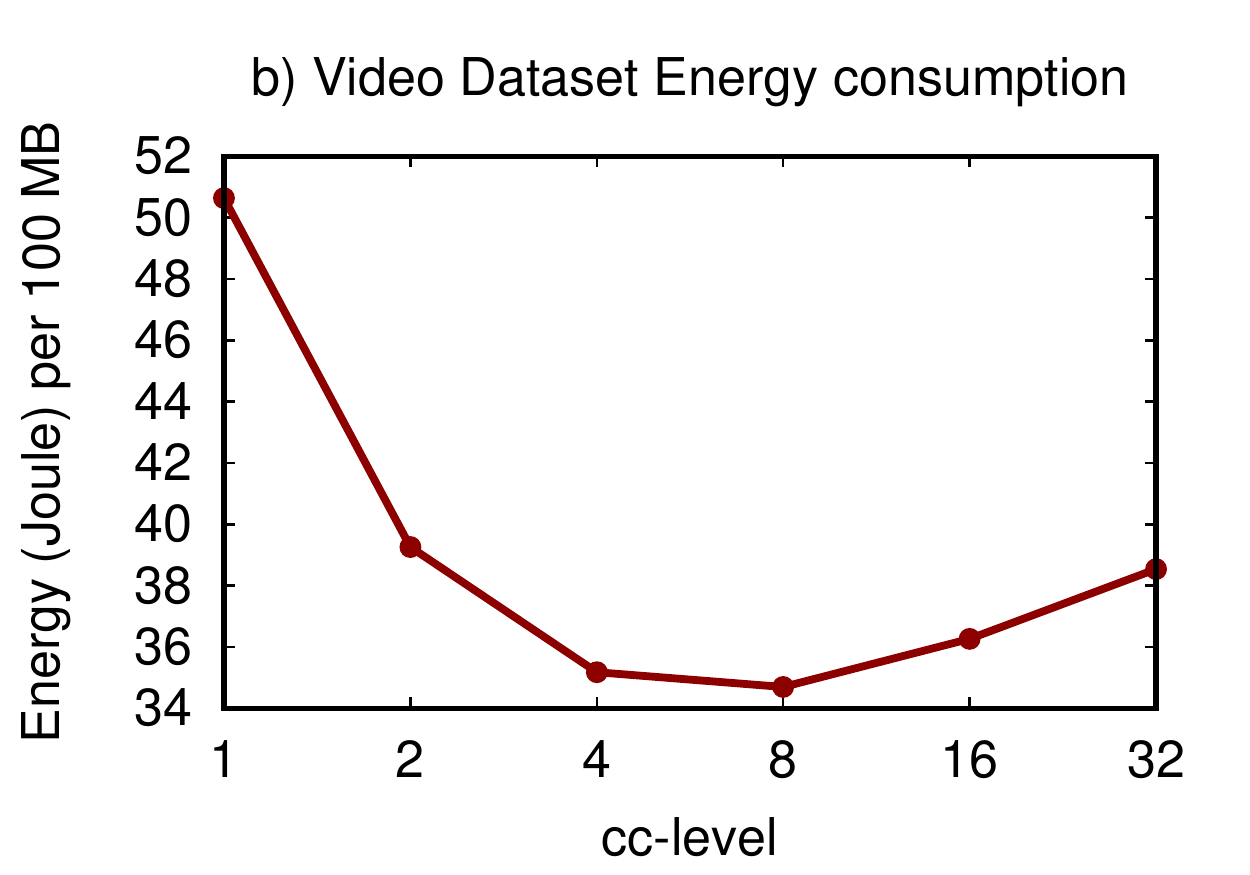}\\
			\includegraphics[keepaspectratio=true,angle=0,width=130mm]{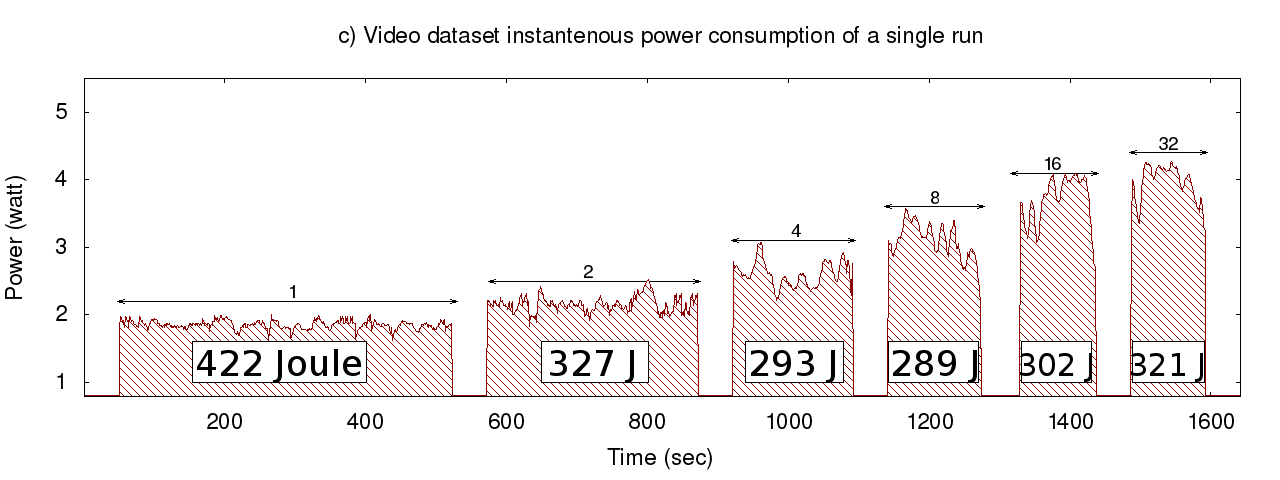}
		\end{tabular}
		\caption{(a) Throughput; (b) Energy Consumption per 100 MB; and (c) Instantaneous Power Consumption of a single video dataset transfer with increased level of concurrency from AWS EC2 Sydney to DIDCLAB Galaxy S5.} \label{fig:instantpower}
	\end{centering}
\end{figure*}

Figure~\ref{fig:phones} presents the results of achieved throughput and energy consumption for different smartphone clients during data transfers from the web server on AWS EC2 Sydney to the clients located at DIDCLAB in Buffalo. We picked four different featured smartphones (as presented in Table~\ref{tab:phonespecs}) in order to test and analyze their performance on three datasets mentioned above. While earlier released smartphones such as Nexus S, Google Nexus N3 adopted mediocre throughput, recently released smartphones such as Galaxy S4, Galaxy S5 attained noticeably better performance. All phones showed improved end-to-end throughput for html, image and video datasets when concurrency level is increased up to a specific level. For html dataset,  concurrency parameter improved throughput for Nexus S by increasing throughput value from 1.8 Mbps to 19 Mbps (10.55X improvement), for Galaxy Nexus N3 by increasing from 1.9 Mbps to 22 Mbps (11X), for Galaxy S4 by increasing from 2.3 Mbps to 36 Mbps (15.6X) and lastly for Galaxy S5 by increasing from 2.6 Mbps to 42 Mbps (16.7X). For the image dataset, throughput increased 1.87X times for Nexus S, 2.4X for Nexus N3, 5.3X for Galaxy S4, and 8.5X for Galaxy S5. Lastly, throughput for video dataset was improved 1.14X times for Nexus S, 1.19X for Nexus N3, 3X for Galaxy S5 and 4.5X for Galaxy S5. Overall, best throughput was gained with Galaxy S5 using the html dataset, with almost 17X increase.

Each smartphone's energy consumption per 100 MB also decreased as we increased concurrency level from 1 to 16 for the html dataset. While further increasing the level of concurrency caused an increase in the energy consumption for Nexus S and Nexus N3, it still continued to decrease for Galaxy S5, which obtained the highest energy saving with 70\% per 100 MB. The throughput of Nexus S and Galaxy Nexus N3 showed logarithmic behavior for the image and video datasets. In fact, concurrency doubled the throughput from 1 to 4 for image dataset and increased throughput by nearly 135\% for video dataset on Nexus S and Galaxy Nexus N3. Additionally, the energy saving rates of Nexus S and Galaxy Nexus N3 for the image dataset from concurrency level 1 to 4 were 33\% (at level 16: 34\%), 29\% (at level 16: 35\%), and  for video dataset 6\% (at level 16: 7\%), 9\% (at level 16: 13\%), respectively. On the other hand, we obtained higher throughput values for image and video datasets on Galaxy S4 and Galaxy S5. While throughput increased up to concurrency level 32, energy saving rate did not decrease as proportionally, namely it started to increase after concurrency level 16 for both S4 and S5. Galaxy S5 succeeded to increase the throughput 8.5X times and 4.5X times for image and video datasets respectively. It also saved 68\% energy for image dataset and 38\% for video dataset. The smartphones with powerful processors, larger memory and optimized OS that is high compatible with all subsystems distinctively separated from earlier released ones. 

We also conducted experiments between three different web servers at different locations and the mobile client at DIDCLAB in Buffalo. 
The RTT between mobile client at DIDCLAB to Chameleon Cloud, AWS EC2 Frankfurt, and AWS ECS Sydney are 59 ms, 115 ms, and 290 ms respectively. The average throughput vs energy consumption trade-offs from those three servers to the client Samsung Galaxy S5  at DIDCLAB can be seen in Figure~\ref{fig:servers}. As shown in Figure~\ref{fig:servers}(a)-(f), increasing the level of concurrency up to 16 improved the end-to-end data transfer throughput while reducing the energy consumption. However, we should emphasize that increasing concurrency does not always improve throughput. Although we have results of concurrency level 64 as well, we did not include it in the figure to protect the overall congruity of graphs presented in the paper. At concurrency level 64, throughput starts to decrease for the image and video datasets while it still continues to slightly increase for the html dataset. 
Although the throughput curves showed similar increasing pattern for all three servers, the highest throughput is achieved from the Chameleon Cloud in Austin node, which is 105 Mbps for the html, 117 Mbps for the image, and 124 Mbps for the video dataset. Energy consumption rates on three servers also showed similarity in decreasing when the concurrency level is changed from 1 to 16, and then it comes to a balance. Overall, we obtained the best energy savings for the html dataset, which was 50\% on Chameleon Cloud, 70\% on EC2 Sydney, and 65\% on EC2 Frankfurt.

Having throughput and energy consumption results of individual parameters on data transfers, we designed a simple download manager that uses the combination of concurrency and parallelism parameters in order to increase throughput while saving energy. Three dimensional Figure~\ref{fig:matlabS5} shows throughput versus energy consumption (per 100 MB) trade-offs of our application level parameters on the html, image, video, and 32 GB datasets from web server at AWS EC2 Sydney to Galaxy S5 at DIDCLAB in Buffalo. As stated earlier~\cite{YildirimPCP}, parallelism is a parameter that is more effective on data transfers consisting of large files. From the combined parameter results of Galaxy S5, we also examined that parallelism becomes effective after reaching specific file size threshold, and improved throughput up to optimal parameter value depending on the dataset. When concurrency level increased from 1 to 32 as well as parallelism from 1 to 8, throughput slightly improved for the html and image datasets compared to individual parameter results in terms of parallelism as seen in Figure~\ref{fig:matlabS5}(a)-(d) and energy consumption per 100 MB increased when the level of parallelism increased on fixed concurrency level. On the other hand, concurrency still managed to show its positive effect on throughput at each parallelism level. Additionally, parallelism became more effective in case of smaller buffer size is used with large Bandwidth Delay Product (BDP), which is a product of bandwidth and RTT, as shown Figure~\ref{fig:matlabS5}(e)-(h) for Video and 32 GB datasets. Since this occurs with longer RTTs, we presented the examined results of AWS EC2 Sydney node with 290 ms RTT. The RTT from Galaxy S5 to Chameleon and EC2 Frankfurt is 59 ms and 115 ms respectively. For the video dataset, throughput increased for both parallelism and concurrency from level 1 to 8, and from 1 to 32 respectively. While energy consumption decreased up to concurrency level 8 and parallelism level 4 (before coming to a balance or increase again) for the video dataset, it decreased up to concurrency level 16 and parallelism level 8 for the 32 GB dataset. Overall, using the combined parameters, we managed to increase the highest energy saving result of individual parameter(70\%) even further (up to 81\%). We run the same experiments with other smartphones as well: while Galaxy S4 presented similar but less throughput and higher energy efficiency, Galaxy Nexus N3 and Nexus S showed moderate performance compared to Galaxy S5.

Performing back-to-back sequential transfers and concurrent transfers have the effect of reducing the ratio of tail energy compared to the total energy consumption during the data transfer as presented in Figure~\ref{fig:instant4Gwifi}. Tail energy is referred to the wasted energy consumption in high-power state of the mobile device after the completion of a data transfer~\cite{balasubramanian2009energy}. 
In this experiment, we first transfer a single file, then measure both the energy during actual file transfer and the tail energy.  We repeat the same experiment for back-to-back sequential transfer of two files and also for concurrent transfer of two files. We run these experiments for both WiFi and 4G LTE. The results are presented in Figure~\ref{fig:instant4Gwifi}. 
As seen in the figure, 4G LTE has a much longer tail time compared to WiFi. This tail time is almost constant for the 4G LTE and WiFi transfers regardless of the size of the file transferred. This means that it has a much larger share for small file transfers and smaller share for the large file transfers. One mechanism to reduce the overall impact of the tail energy is to transfer the files back-to-back without leaving any space between individual file transfers. This way, we only pay for the tail energy once instead of paying for it for every file transfer. 
Concurrency has exact the same effect as back-to-back transfers in reducing the impact of tail energy as seen in the figure.

Figure~\ref{fig:instantpower} shows the achieved end-to-end throughput, total energy consumption, and the change in instantaneous power consumption during one of the video dataset transfers from AWS EC2 Sydney to the client Galaxy S5 at DIDCLAB with increased concurrency level. 
This figure presents the break point for the throughput versus energy consumption trade-off very well. 
As long as the energy gain due to the decreased transfer time is more than the loss due to the
increased instantaneous power consumption, then we save energy at this device while
increasing the throughput. But this is not always the case as seen in Figure~\ref{fig:instantpower}(c). We observe that although the throughput continues to increase after concurrency level 8, the total energy consumption does not continue to decrease, instead comes to a balance and starts increasing again.
The main reason for this is the server and network components are typically not energy proportional.




All these experiments taught us the following {\bf lessons} that could be used to generate energy-aware models or algorithms for data transfers with mobile devices:

\begin{itemize}

\item Using concurrency is highly beneficial and helpful in terms of getting increased throughput as well as decreased total energy consumption up to the optimal concurrency level depending on the dataset characteristics, the resource and capabilities of the mobile device, and network conditions. Additionally, we learned that increasing concurrency beyond optimal point is not always effective in terms of achieving higher throughput and better energy efficiency. 

\item Concurrency also has the effect of reducing the impact of tail energy (the wasted energy consumption in high-power state of the mobile device) after the completion of the data transfers.

\item Parallelism is advantageous when it is used for the transfer of large files with respect to the Bandwidth-Delay-Product and TCP buffer size. 

\item I/O request size on the mobile device contributes to the end-to-end data transfer throughput if the client storage I/O is the main bottleneck in the system.

\item Finding optimal parameter values for these parameters is a challenging task because incorrect tuning of the parameters can cause underutilization of the network and unnecessarily high energy consumption. 

\item With optimal individual parameters, up to 70\% energy saving can be attained on the mobile device during HTTP data transfers while increasing the throughput at the same time.

\item Using combined parameters with optimal individual values boosts the throughput as well as energy saving, in particular for large data transfers. With optimal parameter combination of concurrency and parallelism parameters, up to 81\% energy saving could be attained on the mobile device during HTTP data transfers.

\item More energy savings can be attained in long RTT networks under same circumstances, but the downside is that the achieved highest throughput value is lower even though the throughput still continues to increase. For example, the achieved energy savings for the same dataset is 50\% for Chameleon Cloud, 65\% for AWS EC2 Frankfurt, and 70\% for AWS EC2 Sydney. On the other hand, the achieved highest throughput value for the exact same dataset is 106 Mbps for Chameleon Cloud, 55 Mbps for AWS EC2 Frankfurt, and 42 Mbps for AWS EC2 Sydney.

\end{itemize}

\section{Conclusion and Future Work}
\label{sec:conclusion}

In this paper, we performed extensive analysis and presented the effects of application-layer data transfer protocol parameters (such as the number of parallel data streams per file, the level of concurrent file transfers to fill the mobile network pipes, and the I/O request size) on mobile data transfer throughput and energy consumption for WiFi and 4G LTE connections. 
Our analysis shows that significant energy savings can be achieved with application-layer solutions at the mobile systems during data transfer with no or minimal performance penalty.
We also show that, in many cases, performance increase and energy savings can be achieved simultaneously.
According to our experiments, by only tuning the concurrency and parallelism levels during data transfers, an energy saving up to 81\% can be achieved during mobile networked I/O. At the same time, the throughput of the end-to-end data transfer can be increased by up to 8.5X. Concurrency also has the effect of reducing the otherwise wasted tail energy, further minimizing the total energy consumption on the mobile device.

As the future work, we plan to develop optimization algorithms/models that will provide the users the ability to dynamically choose the optimal combination of transfer protocol parameters for high throughput performance as well as increased energy efficiency. 

\bibliographystyle{plain}
\bibliography{main}

\begin{thebibliography}{10}

\bibitem{akella2001protocols}
Srinivasa~Aditya Akella, Rajesh~Krishna Balan, and Nikhil Bansal.
\newblock Protocols for low-power.
\newblock 2001.

\bibitem{Kosar_jrnl14}
I.~Alan, E.~Arslan, and T.~Kosar.
\newblock {Energy-Performance Trade-offs in Data Transfer Tuning at the
  End-Systems}.
\newblock {\em Sustainable Computing: Informatics and Systems Journal},
  4:4:318-329, 2014.

\bibitem{Alan2015}
Ismail Alan, Engin Arslan, and Tevfik Kosar.
\newblock Power-aware data scheduling algorithms.
\newblock In {\em Proceedings of IEEE/ACM Supercomputing Conference (SC'15)},
  November 2015.

\bibitem{globusonline}
B.~Allen, J.~Bresnahan, L.~Childers, I.~Foster, G.~Kandaswamy, R.~Kettimuthu,
  J.~Kordas, M.~Link, S.~Martin, K.~Pickett, and S.~Tuecke.
\newblock Software as a service for data scientists.
\newblock {\em Communications of the {ACM}}, 55:2:81--88, 2012.

\bibitem{R_Balak98}
H.~Balakrishman, V.~N. Padmanabhan, S.~Seshan, M.~Stemm, and R.~H. Katz.
\newblock Tcp behavior of a busy internet server: Analysis and improvements.
\newblock In {\em Proceedings of INFOCOM '98}, pages 252--262. IEEE, March
  1998.

\bibitem{balasubramanian2009energy}
Niranjan Balasubramanian, Aruna Balasubramanian, and Arun Venkataramani.
\newblock Energy consumption in mobile phones: a measurement study and
  implications for network applications.
\newblock In {\em Proceedings of the 9th ACM SIGCOMM conference on Internet
  measurement conference}, pages 280--293. ACM, 2009.

\bibitem{bertozzi2002power}
Davide Bertozzi, Luca Benini, and Bruno Ricco.
\newblock Power aware network interface management for streaming multimedia.
\newblock In {\em Wireless Communications and Networking Conference, 2002.
  WCNC2002. 2002 IEEE}, volume~2, pages 926--930. IEEE, 2002.

\bibitem{bertozzi2003transport}
Davide Bertozzi, Anand Raghunathan, Luca Benini, and Srivaths Ravi.
\newblock Transport protocol optimization for energy efficient wireless
  embedded systems.
\newblock In {\em Proceedings of the conference on Design, Automation and Test
  in Europe-Volume 1}, page 10706. IEEE Computer Society, 2003.

\bibitem{bharghavan1994macaw}
Vaduvur Bharghavan, Alan Demers, Scott Shenker, and Lixia Zhang.
\newblock Macaw: a media access protocol for wireless lan's.
\newblock {\em ACM SIGCOMM Computer Communication Review}, 24(4):212--225,
  1994.

\bibitem{carroll2010analysis}
Aaron Carroll and Gernot Heiser.
\newblock An analysis of power consumption in a smartphone.
\newblock In {\em USENIX annual technical conference}, volume~14. Boston, MA,
  2010.

\bibitem{chandra2002application}
Surendar Chandra and Amin Vahdat.
\newblock Application-specific network management for energy-aware streaming of
  popular multimedia formats.
\newblock In {\em USENIX Annual Technical Conference, General Track}, pages
  329--342, 2002.

\bibitem{chang2000energy}
Jae-Hwan Chang and Leandros Tassiulas.
\newblock Energy conserving routing in wireless ad-hoc networks.
\newblock In {\em INFOCOM 2000. Nineteenth Annual Joint Conference of the IEEE
  Computer and Communications Societies. Proceedings. IEEE}, volume~1, pages
  22--31. IEEE, 2000.

\bibitem{cianca2001improving}
E~Cianca, M~Ruggieri, and R~Prasad.
\newblock Improving tcp/ip performance over cdma wireless links: A physical
  layer approach.
\newblock In {\em Personal, Indoor and Mobile Radio Communications, 2001 12th
  IEEE International Symposium on}, volume~1, pages A--83. IEEE, 2001.

\bibitem{correia2010challenges}
Luis~M Correia, Dietrich Zeller, Oliver Blume, Dieter Ferling, Ylva Jading,
  Istv{\'a}n G{\'o}dor, Gunther Auer, and Liesbet Van Der~Perre.
\newblock Challenges and enabling technologies for energy aware mobile radio
  networks.
\newblock {\em Communications Magazine, IEEE}, 48(11):66--72, 2010.

\bibitem{cui2004energy}
Shuguang Cui, Andrea~J Goldsmith, and Ahmad Bahai.
\newblock Energy-efficiency of mimo and cooperative mimo techniques in sensor
  networks.
\newblock {\em IEEE Journal on selected areas in communications},
  22(6):1089--1098, 2004.

\bibitem{Czyz2014}
Jakub Czyz, Mark Allman, Jing Zhang, Scott IekelJohnson, Eric Osterweil, and
  Michael Bailey.
\newblock Measuring ipv6 adoption.
\newblock {\em SIGCOMM Comput. Commun. Rev.}, 44(4):87--98, August 2014.

\bibitem{specOverview}
Kaivalya~M Dixit.
\newblock Overview of the spec benchmarks., 1993.

\bibitem{Dogar2010}
Fahad~R. Dogar and Peter Steenkiste.
\newblock Catnap: Exploiting high bandwidth wireless interfaces to save energy
  for mobile devices.
\newblock In {\em Proc. Int. Conf. Mobile Systems, Applications and Services
  (MobiSys)}, 2010.

\bibitem{edwards2016gamification}
Elizabeth~Ann Edwards, J~Lumsden, C~Rivas, L~Steed, LA~Edwards, A~Thiyagarajan,
  R~Sohanpal, H~Caton, CJ~Griffiths, MR~Munaf{\`o}, et~al.
\newblock Gamification for health promotion: systematic review of behaviour
  change techniques in smartphone apps.
\newblock {\em BMJ open}, 6(10):e012447, 2016.

\bibitem{R_Eggert00}
L.~Eggert, J.~Heidemann, and J.~Touch.
\newblock Effects of ensemble-tcp.
\newblock {\em ACM SIGCOMM Computer Communication Review}, 30(1):15--29,
  January 2000.

\bibitem{haas1997mobile}
Zygmunt~J Haas.
\newblock Mobile-tcp: an asymmetric transport protocol design for mobile
  systems.
\newblock In {\em Mobile Multimedia Communications}, pages 117--128. Springer,
  1997.

\bibitem{R_Hacker02}
T.~J. Hacker, B.~D. Noble, and B.~D. Atley.
\newblock The end-to-end performance effects of parallel tcp sockets on a lossy
  wide area network.
\newblock In {\em Proceedings of IPDPS '02}, page 314. IEEE, April 2002.

\bibitem{R_Hacker05}
T.~J. Hacker, B.~D. Noble, and B.~D. Atley.
\newblock Adaptive data block scheduling for parallel streams.
\newblock In {\em Proceedings of HPDC '05}, pages 265--275. ACM/IEEE, July
  2005.

\bibitem{R_Karrer06}
R.~P. Karrer, J.~Park, and J.~Kim.
\newblock Tcp-rome:performance and fairness in parallel downloads for web and
  real time multimedia streaming applications.
\newblock In {\em Technical Report}. Deutsche Telekom Laboratories, September
  2006.

\bibitem{kellerman2016daily}
Aharon Kellerman.
\newblock {\em Daily spatial mobilities: Physical and virtual}.
\newblock Routledge, 2016.

\bibitem{khan2006application}
Shoaib Khan, Yang Peng, Eckehard Steinbach, Marco Sgroi, and Wolfgang Kellerer.
\newblock Application-driven cross-layer optimization for video streaming over
  wireless networks.
\newblock {\em IEEE Communications Magazine}, 44(1):122--130, 2006.

\bibitem{DISCS12}
J.~Kim, E.~Yildirim, and T.~Kosar.
\newblock A highly-accurate and low-overhead prediction model for transfer
  throughput optimization.
\newblock In {\em Proc. of {DISCS} Workshop}, November 2012.

\bibitem{Cluster_2015}
JangYoung Kim, Esma Yildirim, and Tevfik Kosar.
\newblock A highly-accurate and low-overhead prediction model for transfer
  throughput optimization.
\newblock {\em Cluster Computing}, 18(1):41--59, 2015.

\bibitem{WORLDS_2004}
George Kola, Tevfik Kosar, Jaime Frey, Miron Livny, Robert Brunner, and Michael
  Remijan.
\newblock Disc: A system for distributed data intensive scientific computing.
\newblock In {\em WORLDS}, 2004.

\bibitem{Kosar09}
T.~Kosar and M.~Balman.
\newblock A new paradigm: Data-aware scheduling in grid computing.
\newblock {\em Future Generation Computing Systems}, 25(4):406--413, 2009.

\bibitem{kosar04}
T.~Kosar and M.~Livny.
\newblock Stork: Making data placement a first class citizen in the grid.
\newblock In {\em Proceedings of ICDCS'04}, pages 342--349, March 2004.

\bibitem{Thesis_2005}
Tevfik Kosar.
\newblock {\em Data Placement in Widely Distributed Sytems}.
\newblock PhD thesis, University of Wisconsin--Madison, 2005.

\bibitem{IGI_2012}
Tevfik Kosar.
\newblock Data intensive distributed computing: Challenges and solutions for
  large-scale information management, 2012.

\bibitem{ScienceCloud_2013}
Tevfik Kosar, Engin Arslan, Brandon Ross, and Bing Zhang.
\newblock Storkcloud: Data transfer scheduling and optimization as a service.
\newblock In {\em Proceedings of the 4th ACM workshop on Scientific cloud
  computing}, pages 29--36. ACM, 2013.

\bibitem{Royal_2011}
Tevfik Kosar, Mehmet Balman, Esma Yildirim, Sivakumar Kulasekaran, and Brandon
  Ross.
\newblock Stork data scheduler: Mitigating the data bottleneck in e-science.
\newblock {\em Philosophical Transactions of the Royal Society of London A:
  Mathematical, Physical and Engineering Sciences}, 369(1949):3254--3267, 2011.

\bibitem{krashinsky2005minimizing}
Ronny Krashinsky and Hari Balakrishnan.
\newblock Minimizing energy for wireless web access with bounded slowdown.
\newblock {\em Wireless Networks}, 11(1-2):135--148, 2005.

\bibitem{kravets2000application}
Robin Kravets and Parameshwaran Krishnan.
\newblock Application-driven power management for mobile communication.
\newblock {\em Wireless Networks}, 6(4):263--277, 2000.

\bibitem{R_Lee01}
J.~Lee, D.~Gunter, B.~Tierney, B.~Allcock, J.~Bester, J.~Bresnahan, and
  S.~Tuecke.
\newblock Applied techniques for high bandwidth data transfers across wide area
  networks.
\newblock In {\em International Conference on Computing in High Energy and
  Nuclear Physics}, April 2001.

\bibitem{R_Liu10}
W.~Liu, B.~Tieman, R.~Kettimuthu, and I.~Foster.
\newblock A data transfer framework for large-scale science experiments.
\newblock In {\em Proc. 3rd International Workshop on Data Intensive
  Distributed Computing (DIDC '10) in conjunction with 19th International
  Symposium on High Performance Distributed Computing (HPDC '10)}, June 2010.

\bibitem{R_Lu05}
D.~Lu, Y.~Qiao, and P.~A. Dinda.
\newblock Characterizing and predicting tcp throughput on the wide area
  network.
\newblock In {\em Proceedings of ICDCS '05}, pages 414--424. IEEE, June 2005.

\bibitem{R_Dinda05}
D.~Lu, Y.~Qiao, P.~A. Dinda, and F.~E. Bustamante.
\newblock Modeling and taming parallel tcp on the wide area network.
\newblock In {\em Proceedings of IPDPS '05}, page 68.2. IEEE, April 2005.

\bibitem{chameleon}
Joe Mambretti, Jim Chen, and Fei Yeh.
\newblock Next generation clouds, the chameleon cloud testbed, and software
  defined networking (sdn).
\newblock In {\em Cloud Computing Research and Innovation (ICCCRI), 2015
  International Conference on}, pages 73--79. IEEE, 2015.

\bibitem{nasipuri2000mac}
Asis Nasipuri, Shengchun Ye, J~You, and Robert~E Hiromoto.
\newblock A mac protocol for mobile ad hoc networks using directional antennas.
\newblock In {\em Wireless Communications and Networking Confernce, 2000. WCNC.
  2000 IEEE}, volume~3, pages 1214--1219. IEEE, 2000.

\bibitem{nika2015energy}
Ana Nika, Yibo Zhu, Ning Ding, Abhilash Jindal, Y~Charlie Hu, Xia Zhou, Ben~Y
  Zhao, and Haitao Zheng.
\newblock Energy and performance of smartphone radio bundling in outdoor
  environments.
\newblock In {\em Proceedings of the 24th International Conference on World
  Wide Web}, pages 809--819. International World Wide Web Conferences Steering
  Committee, 2015.

\bibitem{pathak2012energy}
Abhinav Pathak, Y~Charlie Hu, and Ming Zhang.
\newblock Where is the energy spent inside my app?: fine grained energy
  accounting on smartphones with eprof.
\newblock In {\em Proceedings of the 7th ACM european conference on Computer
  Systems}, pages 29--42. ACM, 2012.

\bibitem{pering2006coolspots}
Trevor Pering, Yuvraj Agarwal, Rajesh Gupta, and Roy Want.
\newblock Coolspots: reducing the power consumption of wireless mobile devices
  with multiple radio interfaces.
\newblock In {\em Proceedings of the 4th international conference on Mobile
  systems, applications and services}, pages 220--232. ACM, 2006.

\bibitem{Mirkovic2015}
Philipp Richter, Nikolaos Chatzis, Georgios Smaragdakis, Anja Feldmann, and
  Walter Willinger.
\newblock Distilling the internet's application mix from packet-sampled
  traffic.
\newblock In Jelena Mirkovic and Yong Liu, editors, {\em Passive and Active
  Measurement}, volume 8995 of {\em Lecture Notes in Computer Science}, pages
  179--192. 2015.

\bibitem{schulman2010bartendr}
Aaron Schulman, Vishnu Navda, Ramachandran Ramjee, Neil Spring, Pralhad
  Deshpande, Calvin Grunewald, Kamal Jain, and Venkata~N Padmanabhan.
\newblock Bartendr: a practical approach to energy-aware cellular data
  scheduling.
\newblock In {\em Proceedings of the sixteenth annual international conference
  on Mobile computing and networking}, pages 85--96. ACM, 2010.

\bibitem{schurgers2001modulation}
Curt Schurgers, Olivier Aberthorne, and Mani Srivastava.
\newblock Modulation scaling for energy aware communication systems.
\newblock In {\em Proceedings of the 2001 international symposium on Low power
  electronics and design}, pages 96--99. ACM, 2001.

\bibitem{seada2004energy}
Karim Seada, Marco Zuniga, Ahmed Helmy, and Bhaskar Krishnamachari.
\newblock Energy-efficient forwarding strategies for geographic routing in
  lossy wireless sensor networks.
\newblock In {\em Proceedings of the 2nd international conference on Embedded
  networked sensor systems}, pages 108--121. ACM, 2004.

\bibitem{singh1998pamas}
Suresh Singh and Cauligi~S Raghavendra.
\newblock PamasÑpower aware multi-access protocol with signalling for ad hoc
  networks.
\newblock {\em ACM SIGCOMM Computer Communication Review}, 28(3):5--26, 1998.

\bibitem{singh1998power}
Suresh Singh, Mike Woo, and Cauligi~S Raghavendra.
\newblock Power-aware routing in mobile ad hoc networks.
\newblock In {\em Proceedings of the 4th annual ACM/IEEE international
  conference on Mobile computing and networking}, pages 181--190. ACM, 1998.

\bibitem{R_Sivakumar00}
H.~Sivakumar, S.~Bailey, and R.~L. Grossman.
\newblock Psockets: The case for application-level network striping fpr data
  intensive applications using high speed wide area networks.
\newblock In {\em Proceedings of SC'00 ACM/IEEE conference on Supercomputing},
  pages 37--es. ACM/IEEE, September 2001.

\bibitem{Cisco_2016}
Cisco Systems.
\newblock Visual networking index: Forecast and methodology, 2015--2020, June
  2016.

\bibitem{takai2001effects}
Mineo Takai, Jay Martin, and Rajive Bagrodia.
\newblock Effects of wireless physical layer modeling in mobile ad hoc
  networks.
\newblock In {\em Proceedings of the 2nd ACM international symposium on Mobile
  ad hoc networking \& computing}, pages 87--94. ACM, 2001.

\bibitem{toh2001maximum}
C-K Toh.
\newblock Maximum battery life routing to support ubiquitous mobile computing
  in wireless ad hoc networks.
\newblock {\em IEEE communications Magazine}, 39(6):138--147, 2001.

\bibitem{vallina2011erdos}
Narseo Vallina-Rodriguez and Jon Crowcroft.
\newblock Erdos: achieving energy savings in mobile os.
\newblock In {\em Proceedings of the sixth international workshop on MobiArch},
  pages 37--42. ACM, 2011.

\bibitem{vallina2013energy}
Narseo Vallina-Rodriguez and Jon Crowcroft.
\newblock Energy management techniques in modern mobile handsets.
\newblock {\em Communications Surveys \& Tutorials, IEEE}, 15(1):179--198,
  2013.

\bibitem{aws}
Jinesh Varia and Sajee Mathew.
\newblock Overview of amazon web services.
\newblock {\em Amazon Web Services}, 2014.

\bibitem{woesner1998power}
Hagen Woesner, J-P Ebert, Morten Schlager, and Adam Wolisz.
\newblock Power-saving mechanisms in emerging standards for wireless lans: The
  mac level perspective.
\newblock {\em IEEE Personal Communications}, 5(3):40--48, 1998.

\bibitem{woo2001transmission}
Alec Woo and David~E Culler.
\newblock A transmission control scheme for media access in sensor networks.
\newblock In {\em Proceedings of the 7th annual international conference on
  Mobile computing and networking}, pages 221--235. ACM, 2001.

\bibitem{xu2003impact}
Rong Xu, Zhiyuan Li, Cheng Wang, and Peifeng Ni.
\newblock Impact of data compression on energy consumption of
  wireless-networked handheld devices.
\newblock In {\em Distributed Computing Systems, 2003. Proceedings. 23rd
  International Conference on}, pages 302--311. IEEE, 2003.

\bibitem{xu2001geography}
Ya~Xu, John Heidemann, and Deborah Estrin.
\newblock Geography-informed energy conservation for ad hoc routing.
\newblock In {\em Proceedings of the 7th annual international conference on
  Mobile computing and networking}, pages 70--84. ACM, 2001.

\bibitem{ye2002energy}
Wei Ye, John Heidemann, and Deborah Estrin.
\newblock An energy-efficient mac protocol for wireless sensor networks.
\newblock In {\em INFOCOM 2002. Twenty-First Annual Joint Conference of the
  IEEE Computer and Communications Societies. Proceedings. IEEE}, volume~3,
  pages 1567--1576. IEEE, 2002.

\bibitem{R_Yildirim11}
E.~Yildirim, D.~Yin, and T.~Kosar.
\newblock Prediction of optimal parallelism level in wide area data transfers.
\newblock {\em {IEEE} Transactions on Parallel and Distributed Systems},
  22(12), 2011.

\bibitem{TCC_2016}
Esma Yildirim, Engin Arslan, Jangyoung Kim, and Tevfik Kosar.
\newblock Application-level optimization of big data transfers through
  pipelining, parallelism and concurrency.
\newblock {\em IEEE Transactions on Cloud Computing}, 4(1):63--75, 2016.

\bibitem{DADC_2008}
Esma Yildirim, Mehmet Balman, and Tevfik Kosar.
\newblock Dynamically tuning level of parallelism in wide area data transfers.
\newblock In {\em Proceedings of the 2008 International Workshop on Data-aware
  Distributed Computing}, DADC '08, pages 39--48, New York, NY, USA, 2008. ACM.

\bibitem{YildirimPCP}
Esma Yildirim, JangYoung Kim, and Tevfik Kosar.
\newblock How gridftp pipelining, parallelism and concurrency work: A guide for
  optimizing large dataset transfers.
\newblock In {\em Proceedings of the 2012 SC Companion: High Performance
  Computing, Networking Storage and Analysis}, SCC '12, pages 506--515,
  Washington, DC, USA, 2012. IEEE Computer Society.

\bibitem{NDM_2011}
Esma Yildirim and Tevfik Kosar.
\newblock Network-aware end-to-end data throughput optimization.
\newblock In {\em Proceedings of the first international workshop on
  Network-aware data management}, pages 21--30. ACM, 2011.

\bibitem{JGrid_2012}
Esma Yildirim and Tevfik Kosar.
\newblock End-to-end data-flow parallelism for throughput optimization in
  high-speed networks.
\newblock {\em Journal of Grid Computing}, pages 1--24, 2012.

\bibitem{DADC_2009}
Esma Yildirim, Dengpan Yin, and Tevfik Kosar.
\newblock Balancing tcp buffer vs parallel streams in application level
  throughput optimization.
\newblock In {\em Proceedings of the second international workshop on
  Data-aware distributed computing}, pages 21--30. ACM, 2009.

\bibitem{R_Yin11}
D.~Yin, E.~Yildirim, and T.~Kosar.
\newblock A data throughput prediction and optimization service for widely
  distributed many-task computing.
\newblock {\em {IEEE} Transactions on Parallel and Distributed Systems}, 22(6),
  2011.

\bibitem{zorzi1999tcp}
Michele Zorzi and Ramesh~R Rao.
\newblock Is tcp energy efficient?
\newblock In {\em Mobile Multimedia Communications, 1999.(MoMuC'99) 1999 IEEE
  International Workshop on}, pages 198--201. IEEE, 1999.

\end{thebibliography}
\end{document}